\documentclass[%
 reprint,
 amsmath,amssymb,
 aps
]{revtex4-2}

\usepackage[separate-uncertainty=true]{siunitx}

\usepackage{graphicx}% Include figure files  [dvipdfmx]
\usepackage{subfig}
\usepackage{dcolumn}% Align table columns on decimal point
\usepackage{bm}% bold math
\usepackage{color}% add hypertext capabilities  [dvipdfmx]
\usepackage{ulem}%allows strikethrough via \sout{}
\usepackage[dvipsnames]{xcolor} %this package provides more colours
\newcommand{\revangfreqnominal}{\ensuremath{\omega_{0}}}
\newcommand{\slippage}{\ensuremath{\eta}}

\newcommand{\momentumnominal}{\ensuremath{p_{0}}}
\newcommand{\numprotons}{N}
\newcommand{\lifetime}{\tau}

\begin{document}

\preprint{APS/123-QED}

\title{Beam Stacking Experiment at a Fixed Field Alternating Gradient Accelerator}

\author{T. Uesugi 
\thanks{A footnote to the article title)}}
\email{uesugi.tomonori.2n@kyoto-u.ac.jp}
\author{Y.~Ishi, Y. Kuriyama, Y. Mori}
 
\affiliation{
 Institute for Integrated Radiation and Nuclear Science, Kyoto University\\
 Kumatori-cho, Sennan-gun, Osaka 590-0494 Japan
}

\author{C. Jolly, D.~J.~Kelliher, J.~-B. Lagrange, A.~P.~Letchford, S. Machida, D.~W.~Posthuma de Boer, C.~T.~Rogers, E.~Yamakawa}
\affiliation{
 STFC ISIS Department\\
 Harwell Campus, Didcot, OX11 0QX United Kingdom\\
}

\author{M. Topp-Mugglestone}
\affiliation{
 John Adams Institute, University of Oxford\\
 Keble Road, Oxford, OX1 3RH, United Kingdom\\
}

\date{\today}

\begin{abstract}
A key challenge in particle accelerators is to achieve high peak intensity. 
Space charge is particularly strong at lower energy such as during injection and typically limits achievable peak intensity. 
The beam stacking technique can overcome this limitation by accumulating a beam at high energy where space charge is weaker.
In beam stacking, a bunch of particles is injected and accelerated to high energy. 
This bunch continues to circulate, while a second and subsequent bunches are accelerated to merge into the first. 
It also allows the user cycle and acceleration cycles to be separated which is often valuable. 
Beam stacking is not possible in a time varying magnetic field, but a fixed field machine such as an Fixed Field Alternating Gradient Accelerator (FFA) does not sweep the magnetic field. 
In this paper, we describe experimental demonstration of beam stacking of two beams at KURNS FFA in Kyoto University. 
The momentum spread and intensity of the beam was analysed by study of the Schottky signal, demonstrating stacking with only a slight increase of momentum spread of the combined beams. 
The intensity of the first beam was, however, significantly reduced. 
RF knock-out is the suspected source of the beam loss. 
\end{abstract}

\maketitle

\section{\label{sec:introduction} Introduction}
Next generation intensity-frontier machines demand increasingly high beam powers.
Some applications of high intensity beams, such as proton drivers for a spallation neutron source or neutrino physics\cite{AAPPSBul, SNS, JPARC} require high average beam power. 
Meanwhile, other applications demand high peak beam power as well as high average beam power.
For example, a proton driver for a muon collider would demand high peak beam power because the collider 
luminosity depends on the number of particles-per-bunch \cite{muoncollider}.

Beam energy and average current determine the average beam power of a particle accelerator.
Average current is the product of particles-per-cycle and the repetition rate. To achieve
a higher average beam power, either the repetition rate or particles-per-cycle must be increased.
Whilst the particles-per-cycle has a fundamental
physics limitation in the form of space-charge, the repetition rate is limited only by the technology of the accelerator. From a beam physics perspective, continuous
acceleration is the easiest way to achieve a high average beam current. However,
applications such as spallation neutron sources demand a pulsed time structure \cite{isisii}.

A beam's space charge tune shift is given by \cite{spacechargetuneshift}
	\label{eqn:spaceChargeTuneShift}
\begin{align*}
    \Delta Q_x &= -\frac{r_p n_t R}{\pi \beta^2\gamma^3}
    \frac{1}{\sigma_x(\sigma_x+\sigma_y) Q_x B_f}\\
    \Delta Q_y &= -\frac{r_p n_t R}{\pi \beta^2\gamma^3}
    \frac{1}{\sigma_y(\sigma_x+\sigma_y) Q_y B_f}\\
    \sigma_x &= \sqrt{\tilde{\beta_x} \varepsilon_x +\left(\tilde{D_x} \left( \delta p/p_0\right )\right)^2}\\
    \sigma_y &= \tilde{\beta_y} \varepsilon_y
\end{align*}
where $Q_x$ and $Q_y$ are the horizontal and vertical tunes respectively, $r_p$ is the classical proton radius, $n_t$ is the total number 
of particles,  $R$ is the average machine radius, 
$\beta = \frac{v}{c}$ is the relativistic velocity, $\gamma$ is the Lorentz 
factor, $\varepsilon_{x,y}$ is the geometrical transverse emittance in the horizontal and 
vertical plane respectively, $\tilde{D_x}$ is the average dispersion function, $\delta p/p_0$ 
is the momentum spread, $\tilde{\beta}_{x,y}$ is the average lattice $\beta$ function in the 
horizontal and vertical plane respectively, and $B_f$ is the bunching factor. $\Delta Q_x$ and $\Delta Q_y$ must both 
be less than 1 to avoid crossing
resonances, which limits the number of particles-per-cycle.
It can be seen that $\beta^2 \gamma^3$
increases during acceleration, so $\Delta Q_x$ and $\Delta Q_y$ decrease rapidly with beam energy. In other words, average beam power is limited by
a beam's properties at injection.

This interpretation assumes beams are injected, accelerated and extracted from the ring one at 
a time. If 
multiple beams could be fully accelerated and combined at a higher energy, then $n_t$ could
be increased for a given tune spread. This process, known as ``beam stacking",
is one way to increase the number of particles-per-cycle; although we must
now distinguish machine cycles $T_\textrm{acc}$ and extraction for a user cycle $T_\textrm{user}$.
The time periods of these are related by $T_\textrm{user}= N_s T_\textrm{acc}$,
where $N_s$ is the number of stacked beams. 
The proposed scheme is to increase the number of particles-per-user-cycle by stacking
multiple beams at the extraction energy. Since the number of particles-per-accelerator-cycle
is smaller by $N_s$ times, the space charge spread need not be a limitation.
To avoid large values of $T_\textrm{user}$, $T_\textrm{acc}$ must be minimised, and
a proper stacking procedure must be established to prevent beam loss and control emittance.

By allowing the extraction rate and intensity to be varied cycle-to-cycle, a spallation
neutron source based on stacking would be a first of its kind in terms of experimental flexibility.
Target stations could be envisaged with different repetition rates and different 
peak beam powers, whilst maintaining the same average beam power. Users of such a facility would 
have the option of lower-intensity rapid pulses, or higher-intensity slow pulses. The 
implications of this new regime are yet to be fully explored.

Fixed Field Alternating Gradient Accelerators (FFAs) use fixed-field magnets whose strengths
vary with a beam's orbit. They differ from synchrotrons, whose magnet strength is synchronised with 
beam momentum~\cite{okawaffa, ffa_d, ffa_e, ffa_b, ffa_c}. The cycle time of an FFA is defined 
entirely by the available RF power and its low-level control. Repetition rates on 
the order of \SI{}{\kHz} have been demonstrated \cite{popffa}, which represents an order-of-magnitude
improvement compared to current rapid cycling synchrotrons \cite{thomason_isis_2019}.

Since Liouville's theorem prohibits the increase of phase space density by adding particles, the 
stacking technique proposed here stacks consecutive beams on the momentum direction. This increases
longitudinal emittance proportionally to the number of stacked beams, whilst maintaining transverse 
emittance. Such a scheme was first proposed and demonstrated in the FFA accelerator at Midwestern
University Research Association (MURA) \cite{BeamStackingRSI} and developed at the CERN ISR \cite{BeamStackingISR}. It requires an
accelerator with a large momentum acceptance such as an FFA, which accommodates the range from
injection to its design maximum momentum. 
Provided the RF frequency sweep range from injection to extraction is within the revolution frequency of an injected beam, an FFA can simultaneously accelerate one beam whilst storing another in a single machine. This suggests that a high intensity FFA, which exceeds the intensity of conventional accelerators, could be possible, and is one reason they are being considered as a driver for a future 
spallation neutron source \cite{isisii}.

Whilst beam stacking has already been demonstrated, the process must be optimised if it is to be
applied at high intensities. In particular, this study aims to minimise the momentum spread of two 
stacked beams at an operating FFA facility and investigate sources of beam-loss during the process.
In this paper, details of the stacking process are introduced as part of a simulation study in 
Section~\ref{sec:simulation}, and experimental results are then presented in Section~\ref{sec:meas}. 
Experiments were carried out on the FFA accelerator at the Institute for Integrated Radiation
and Nuclear Science, Kyoto University (KURNS), which is introduced in Section~\ref{sec:KURNS FFA}, and
the particulars of these experiments are described in Section~\ref{sec:expt_setup}. The paper
concludes with a discussion and summary in Sections~\ref{sec:discussion}~and~\ref{sec:conclusion}.

\section{\label{sec:KURNS FFA} KURNS FFA}
The beam stacking experiment was performed using the machine development time of the FFA accelerator at KURNS.
The FFA accelerator complex at KURNS was designed as a proton driver for an accelerator driven system~\cite{pyeon}.
It consists of a negative hydrogen ion souce, an \SI{11}{\MeV} linac, a \SI{150}{\MeV} FFA 
accelerator, referred to as the Main Ring (MR), and beam transport lines \cite{KURRIFFA}.
The main parameters of the complex are shown in Table~\ref{tab:complex_param}.

\begin{table}[htb]
  \setlength\tabcolsep{5.5pt}
  \centering
  \caption{Basic Parameters of KURNS FFA Accelerator Complex}
  \label{tab:complex_param}
\begin{ruledtabular}
  \begin{tabular}{lc}
    \multicolumn{2}{c}{Linac} \\
\colrule
    Repetition rate & $<$ \SI{200}{Hz} \\
    Peak current & $<$ \SI{5}{\micro A} \\
    Pulse length & $<$ \SI{100}{\micro s} \\
    Energy & \SI{11}{MeV} \\
\colrule
\colrule
    \multicolumn{2}{c}{MAIN RING} \\
\colrule
    Symmetry & 12 \\
    Field index $k$ & 7.5 \\
    Energy & 11 - \SI{150}{MeV}  \\
    Betatron tune & (3.64, 1.36) \\
    Momentum compaction & 0.12 \\
    Revolution frequency & 1.6 - \SI{4.3}{MHz} \\
    RF voltage & \SI{4}{kV} \\
  \end{tabular}
\end{ruledtabular}
\end{table}

\begin{figure}[b]
    \centering
    \includegraphics[clip,width=0.4\textwidth]{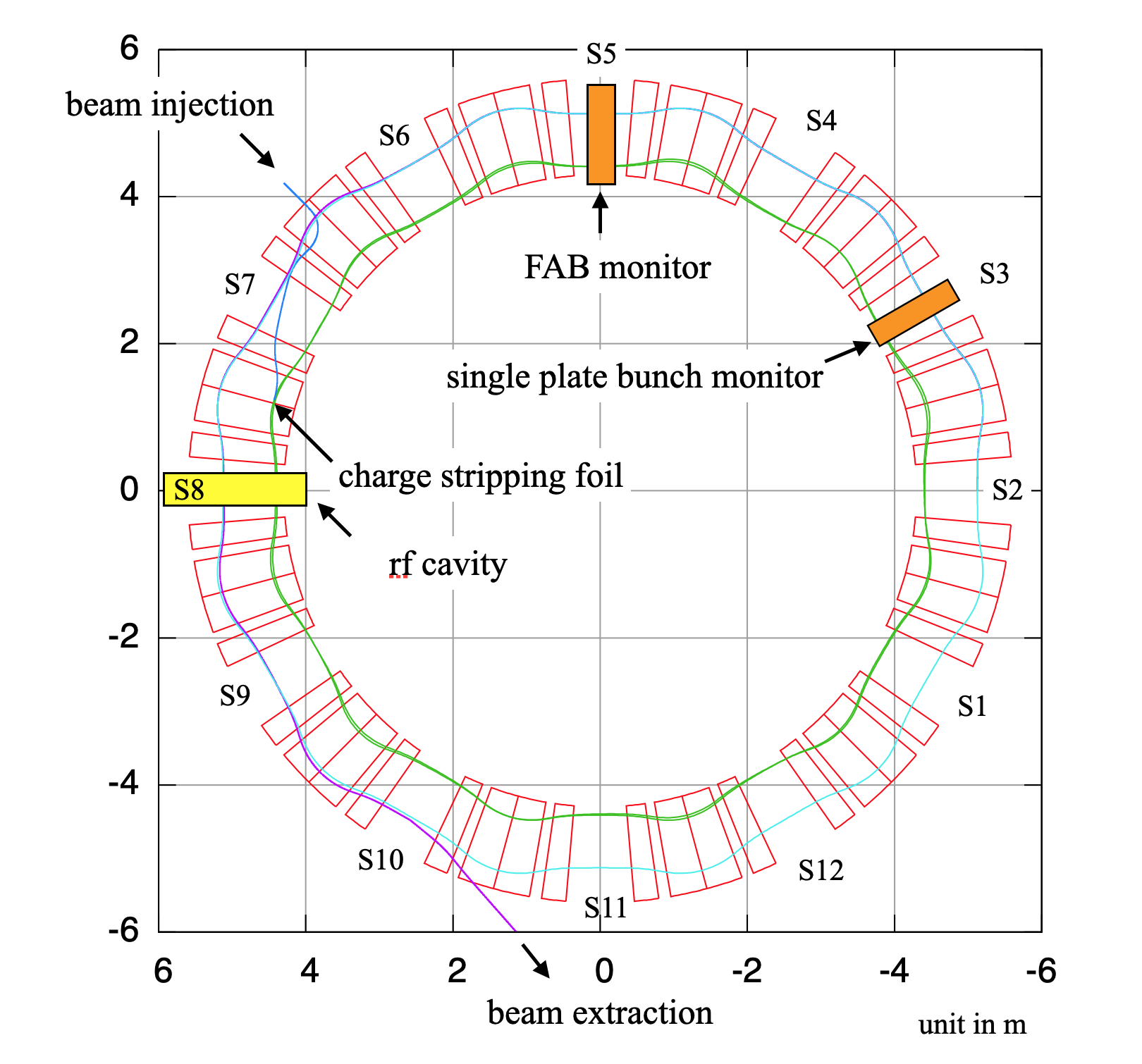}
    \caption{Footprint of KURNS MR. Main magnets are shown as a 
    red rectangular box. Injection and extraction orbits are shown with green and blue lines, 
    respectively. Some equipments are also shown as indicated.}
    \label{fig:footprint}
\end{figure}

The MR is a radial scaling FFA accelerator~\cite{okawaffa, ffa_d, ffa_e} using DFD triplet cells with 12-fold symmetry,
as shown in Fig.~\Ref{fig:footprint}.
Negative hydrogen beams are injected into the MR through a charge stripping foil made of
carbon with thickness of \SI{20}{\micro g/cm^2}. No pulsed magnets are used for the beam injection.
Injected beams are captured and accelerated by a moving RF bucket. In typical operation,
the RF voltage is fixed at \SI{4}{kV} and the synchronous phase is 20~deg over the 
acceleration; the orbit shift at injection corresponds to 1~mm/30~turns (or 24~mm/MeV).
Simulations show that the RF bucket is fully filled in this injection scheme~\cite{tom_injection}.
The revolution time at injection in the MR is \SI{660}{ns}, 
and the typical pulse duration of the injected beam is \SI{100}{\micro s},
which corresponds to about 150 turns.
Two kicker magnets and one septum magnet are used for the beam extraction.

The major instruments used in this experiment are an RF cavity in long straight S8, a full aperture bunch monitor (FAB) in S5, which is discussed in Section~\ref{method:bunchmonitor}, and a position-sensitive single plate bunch monitor in S3.

\section{\label{sec:simulation} Stacking Process in Simulation}
    \begin{figure}[h]
	\centering
	\subfloat[][]{
	\includegraphics[width=0.48\linewidth]{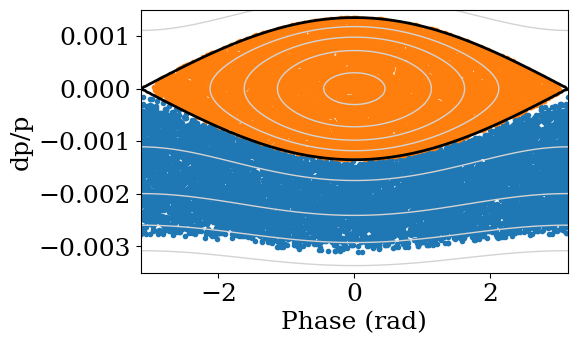}\hfill
	\label{fig:sim:phasesp_a}
	}
	\subfloat[][]{
	\includegraphics[width=0.48\linewidth]{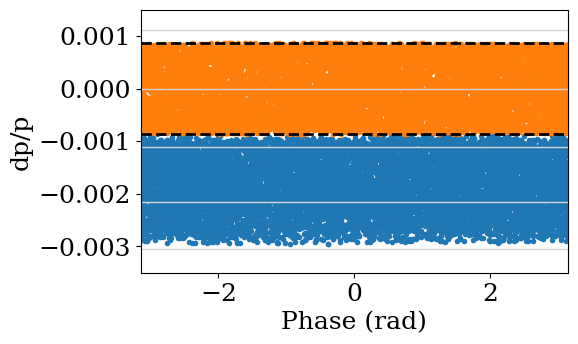}
	\label{fig:sim:phasesp_b}
	}\\
	\subfloat[][]{
	\includegraphics[width=0.48\linewidth]{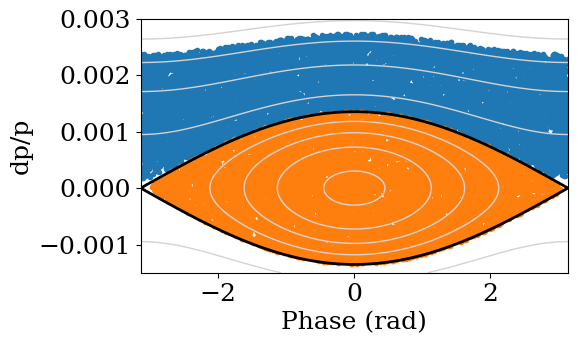}\hfill
	\label{fig:sim:phasesp_c}
}
	\subfloat[][]{
	\includegraphics[width=0.48\linewidth]{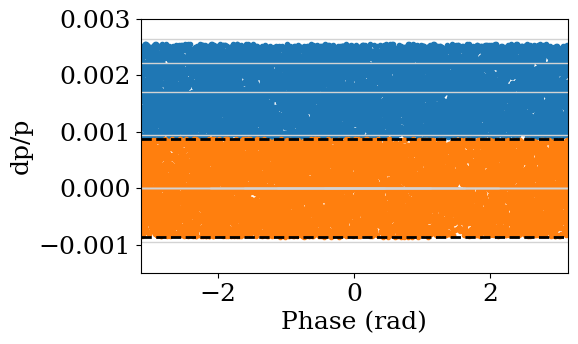}
	\label{fig:sim:phasesp_d}
}
\caption{Longitudinal phase space for the case of 
	 stacking at the top (a,b) and stacking at the bottom (c,d) without gap. The figures on the left (a,c) show the distribution at the point when the synchronous phase reaches zero in each case. The black contour shows the RF bucket separatrix. The figures on the right (b,d) show the distribution after the RF voltage has been gradually reduced to zero. The area between the dashed black lines on the right is equal to the bucket area in the figures on the left.}
\label{fig:sim:idealstacking}
\end{figure}

The aim of the stacking process is to place individual coasting beams next to one another in momentum space, ideally without an intervening gap and without emittance increase. The first step is to accelerate the first bunch to the desired energy $E_{acc,1}=E_{stack}$, with longitudinal emittance $\varepsilon_L$. The RF waveform should ensure that the synchronous phase $\phi_s$ reaches zero once the beam reaches $E_{stack}$. The RF voltage is then reduced sufficiently slowly to achieve adiabatic debunching, i.e. to ensure no emittance increase. In that case, the energy spread of the coasting beam will be $\delta E_b = \varepsilon_L \omega_\textrm{stack}/2\pi$ where $\omega_\textrm{stack}$ is the angular revolution frequency at $E_\textrm{stack}$.

If the second bunch is then accelerated to the same energy $E_{acc,2}=E_\textrm{stack}$, the RF bucket passing through the first beam will result in phase displacement of that beam \cite{PhaseDisplacement}. As a consequence of Liouville's theorem, the final mean energy of the first beam decreases by $\left<\Delta E\right>_D$, 
i.e. $E_{acc,1}=E_{stack}-\left<\Delta E\right>_D$,
and $\left<\Delta E\right>_D=A \omega_\textrm{stack}/2\pi$ where $A$ is the bucket area. To ensure the first and second beams are stacked without separation, $\left<\Delta E\right>_D$ must be equal to the energy spread of each coasting beam $\delta E_b$. It follows that the condition $A=\varepsilon_L$ must be satisfied, i.e. the bucket must be full when the second beam arrives at $E_{stack}$. Since the second beam ends up supplanting the first beam in longitudinal phase space, this case is sometimes called \textit{stacking at the top}. Simulation results depicting this scenario are shown in Fig.\,\ref{fig:sim:phasesp_a},\,\ref{fig:sim:phasesp_b}.

Phase displacement also results in an increase in the energy spread of the first beam once the bucket has passed through. This is because the energy displacement of individual particles depends on their proximity to the RF bucket as they pass the through the out-of-bucket channel\,\cite{Symon_Sessler}. The rms energy increase depends on the area of the RF bucket and the synchronous phase \cite{Jones}. This effect is called scattering. In the simulation results shown in Fig.\,\ref{fig:sim:phasesp_b}, the effect of scattering can be seen in the increased momentum spread of the first beam (blue points) compared with the second beam (orange points). 

In order to avoid the effective emittance increase caused by scattering, the second beam should be placed below the first in momentum space (so-called \textit{stacking at the bottom}). In this case, the RF programme is adjusted so that the second beam arrives just below the first beam in momentum space. Ideally, there should be no gap between the two beams and the second beam should completely fill the stationary bucket (Fig.\,\ref{fig:sim:phasesp_c}). Following adiabatic debunching, an ideal stack of both beams is obtained (Fig.\,\ref{fig:sim:phasesp_d}).

\begin{figure}[h]
	\centering
	\subfloat[][]{
	\includegraphics[width=0.48\linewidth]{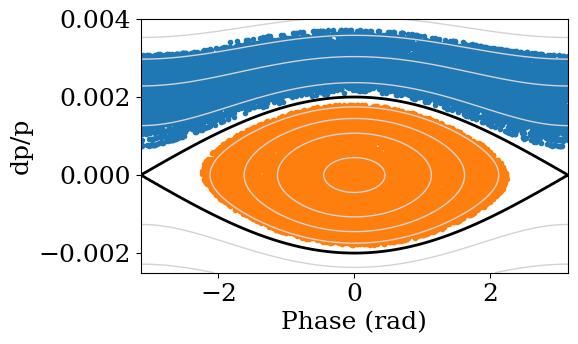}\hfill
	\label{fig:sim:phasesp_exp:subfig1}
	}
	\subfloat[][]{
	\includegraphics[width=0.48\linewidth]{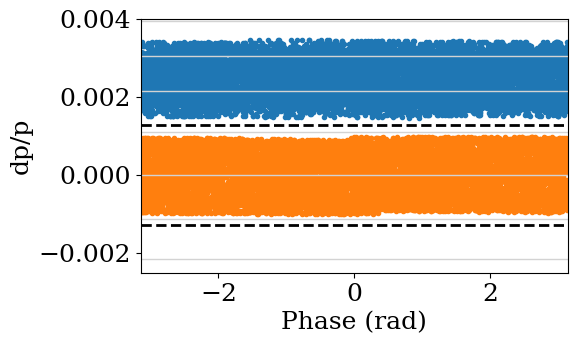}
	\label{fig:sim:phasesp_exp:subfig2}
	}
 \caption{Stacking at the bottom (as in Fig.\,\ref{fig:sim:phasesp_c}, Fig\,\ref{fig:sim:phasesp_d}) but in this case $A > \varepsilon_L$ and the separation between the beams is greater than $\delta E_b$.}
\label{fig:sim:phasesp_exp}
\end{figure}

However, in practice, it may not be possible to stack at the bottom without gap. Here, we describe two considerations that can result in a gap between the two beams. Firstly, if the moving RF bucket makes too close an approach to the already coasting beam, some particles may be phase displaced, resulting in an increase in energy spread. In this case, the separation between the two beams should be greater than $\delta E_b$ to avoid phase displacement. Secondly, it may not be possible for the second beam to completely fill the RF bucket without some particles near the separatrix being lost during acceleration. Fig.~\ref{fig:sim:phasesp_exp} shows a case where both of these conditions apply. The separation between the beams after debunching, shown in Fig.~\ref{fig:sim:phasesp_exp:subfig2}, is a result of the fact the energy separation is greater than $\delta E_b$ and the RF bucket is not filled $(A > \varepsilon_L)$.
\begin{figure}[h]
	\centering
	\subfloat[][]{
	\includegraphics[width=0.48\linewidth]{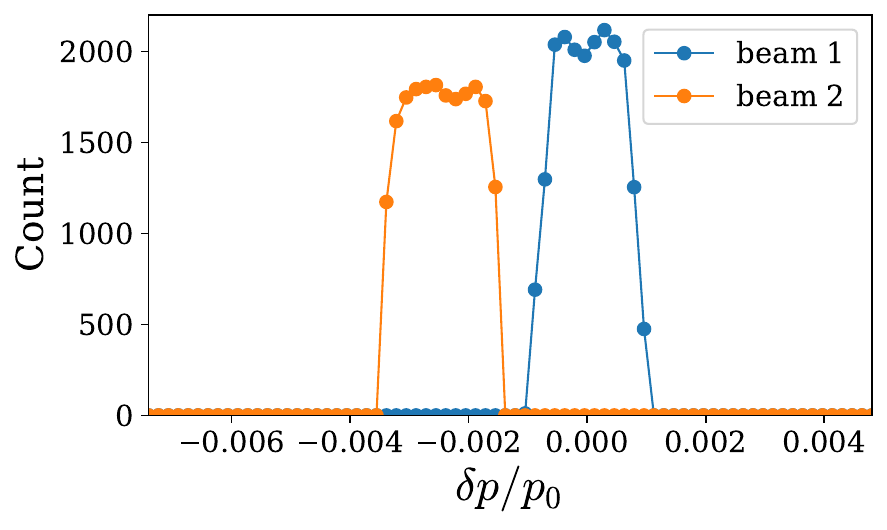}\hfill
	\label{fig:sim:hist_a}
	}
	\subfloat[][]{
	\includegraphics[width=0.48\linewidth]{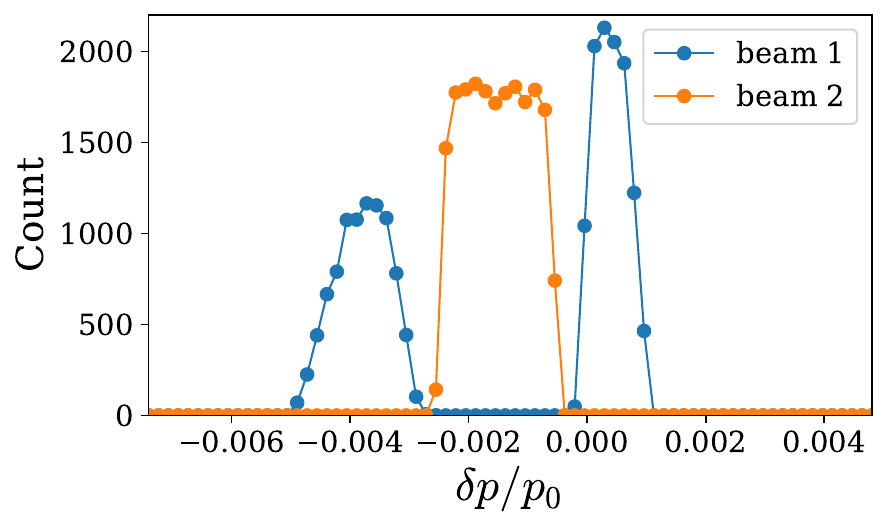}
	\label{fig:sim:hist_b}
	}\\
	\subfloat[][]{
	\includegraphics[width=0.48\linewidth]{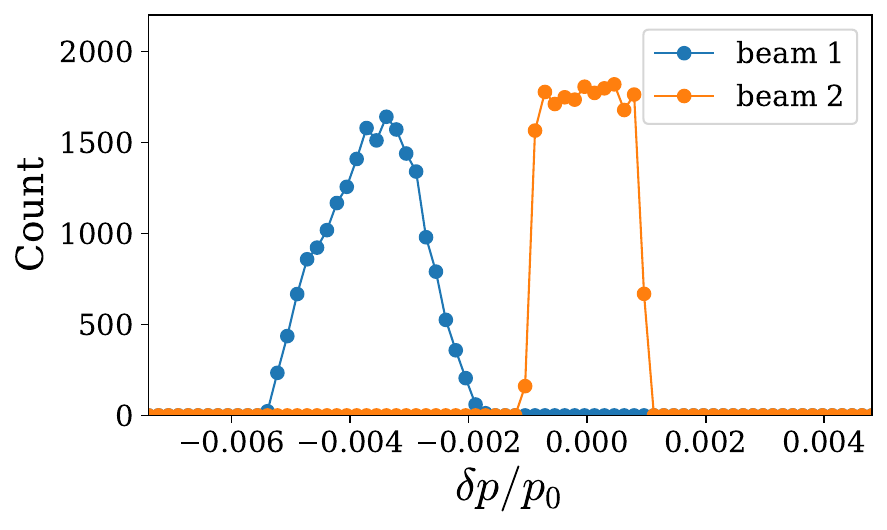}\hfill
	\label{fig:sim:hist_c}
}
	\subfloat[][]{
	\includegraphics[width=0.48\linewidth]{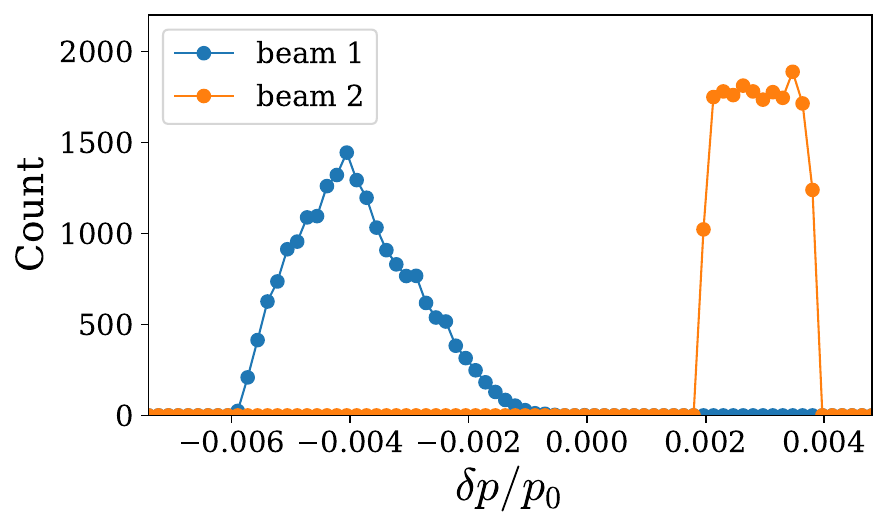}
	\label{fig:sim:hist_d}
}
\caption{Relative momentum projection of two coasting beam distributions when the second beam (orange) is accelerated to a final energy offset $\Delta E_{sep}$ of a) \SI{-168}{\keV}, b) \SI{-103}{\keV}, c) \SI{0}{\keV} and
	d) \SI{200}{\keV} compared to the initial energy of the first beam (blue).}
\label{fig:sim:hist}
\end{figure}

A set of simulations, using a longitudinal tracking code, was carried out in order to calculate the 
optimal conditions for stacking within constraints dictated by our experimental setup. During the 
simulation, the second beam is accelerated and stacked alongside the first beam, which is already 
coasting at energy $E_{stack}$. At the start of each simulation the RF voltage is \SI{4}{\kV} and 
the synchronous phase,  $\phi_s = 20~\textrm{deg}$, according to the nominal operating parameters of the KURNS 
machine. During the first $10$k turns of the simulation, $\phi_s$ is reduced linearly to zero. At 
the same time, the RF voltage is reduced to \SI{0.35}{\kV} in order to reduce the bucket area, and 
hence increase the bucket fill factor (the final bucket fill factor is 0.77). During the subsequent 
$5$k turns, the RF voltage is reduced to zero to debunch the beam. The results of one the 
simulations are shown in Fig.\,\ref{fig:sim:phasesp_exp}.

The longitudinal tracking code integrates the energy $E_i$ and phase $\phi_i$ of each particle $i$ from turn $n$ to $n+1$ with the following \textit{kick-drift} algorithm
\begin{align*}
E_i^{(n+1)} &= E_i^{(n)} + V_0^{(n)} \textrm{sin} \phi_i^{(n)}, \\
\phi_i^{(n+1)} &= \phi_i^{(n)} + 2 \pi \left( \frac{\omega_{RF}(\phi_i^{(n)}) - h 
\omega_i^{(n+1)}}{\omega_i^{(n+1)}}\right),
\label{integration}
\end{align*}
where $V_0^{(n)}$ is RF voltage amplitude at turn $n$ and the angular revolution frequency of each 
particle $\omega_i$ follows from the FFA scaling condition~\cite{okawaffa, ffa_d, ffa_e}
\begin{align*}
\omega_i = \omega_\textrm{ref}\frac{E_\textrm{ref}}{E_i}\left(\frac{p_i}{p_\textrm{ref}}\right)^{\frac{k}{k+1}},
\end{align*}
where $p_i$ is the momentum of each particle, $\omega_\textrm{ref}$ and $E_\textrm{ref}$ are the angular frequency and the total energy at reference momentum $p_\textrm{ref}$ respectively
and $k$ is the field index.
The RF frequency term  $\omega_{RF}(\phi_i^{n})$ assumes a linear variation in RF frequency between 
$\omega_{RF}^n$ to $\omega_{RF}^{n+1}$ depending on the phase of each particle $\phi_i^{(n)}$. It 
should be 
noted that this algorithm implicitly includes phase slip to all orders.  High order terms in phase 
slip need to be included for accuracy when the difference between the RF frequency and the 
revolution frequency of the coasting beam is significant. 

\begin{figure}[h]
	\centering
	\includegraphics[width=0.95\linewidth]{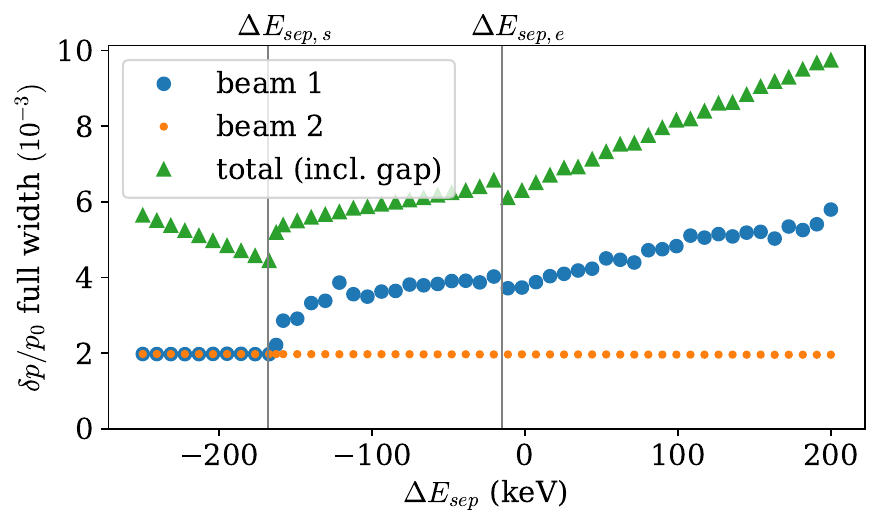}
	\caption{Relative momentum width as a function of $\Delta E_{sep}$. The green points show the full 
	width, including any separation between the beams. The blue and orange points show the full 
	width of the first and second beams respectively. In cases where the first beam is split, as 
	in Fig.\,\ref{fig:sim:hist}(b), the blue points show the sum of the two 
	widths. The vertical lines divide the plot into the three regimes described in the text, from 
	the left: that of no crossing, partial crossing or complete crossing of the first beam by the 
	RF bucket.}
	\label{fig:sim:scan}
\end{figure}

For each simulation, the final energy of the second beam is set to be greater than or less than the initial energy of the first coasting beam $E_{stack}$ (we refer to the difference as the energy separation $\Delta E_{sep}=E_{acc,2}-E_{stack}$, where $E_{acc,2}$ is the final energy of the second beam). This is done by adjusting the initial energy of the second beam at the start of the simulation but keeping the voltage and $\phi_s$ pattern fixed.  The momentum projections shown in Fig.\,\ref{fig:sim:hist} show some representative examples of the coasting beam distribution, depending on the choice of $\Delta E_{sep}$.  The full momentum spread spanning both beams, as well as the momentum spread of the individual beams, is shown as a function of $\Delta E_{sep}$ in Fig.\,\ref{fig:sim:scan}.  

We can identify three regimes in Fig.\ref{fig:sim:scan} ,corresponding to the case where the RF bucket doesn't cross the first beam  ($\Delta E_{sep} \leq \Delta E_{sep,s}$), the partial crossing regime where just part of the first beam is phase displaced ($\Delta E_{sep,s} < \Delta E_{sep} < \Delta E_{sep,e}$), and the complete crossing regime $\Delta E_{sep} \ge \Delta E_{sep,e}$. In the no crossing case, there is no change in the momentum spread of either beam. On the other hand, in the partial or complete crossing cases ($\Delta E_{sep} > \Delta E_{sep},s$), the momentum spread of the first beam tends to increase with $\Delta E_{sep}$. This is because both $\phi_s$ and $V_0$ at the point of RF bucket crossing, and hence the scattering resulting from phase displacement, increases with $\Delta E_{sep}$.  In the complete crossing case there is also an increase in mean phase displacement $\left<\Delta E\right>_D$ with $\Delta E_{sep}$ caused by the increase of bucket area at the point of crossing. This contributes to the increase in the total momentum spread with $\Delta E_{sep}$. 

The minimum total momentum spread in Fig.\,\ref{fig:sim:scan} is at energy separation $\Delta E_{sep,s}$. This is the case of stacking at the bottom, where the RF bucket makes its closest approach to the first beam without causing phase displacement. The corresponding phase space distribution and momentum projection are shown in  Fig.\,\ref{fig:sim:phasesp_exp} and Fig.\,\ref{fig:sim:hist_a}, respectively. 

\section{Experimental setup} \label{sec:expt_setup}

	\subsection{\label{sec:expt_setup:awg} Acceleration Pattern by AWG}
\begin{figure*}
\centering
\subfloat[Injection, acceleration and stacking of the two beams.]{
\includegraphics[width=\columnwidth]{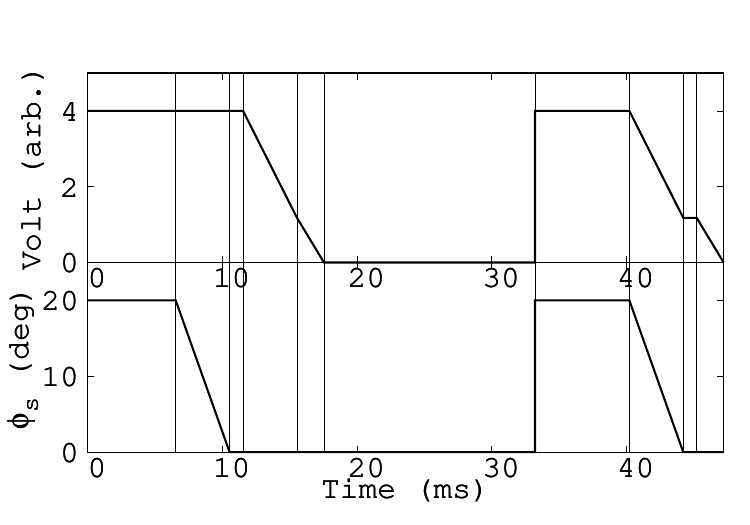}
}
\hspace{0mm}
\subfloat[Recapture of the two beams.]{
\includegraphics[width=\columnwidth]{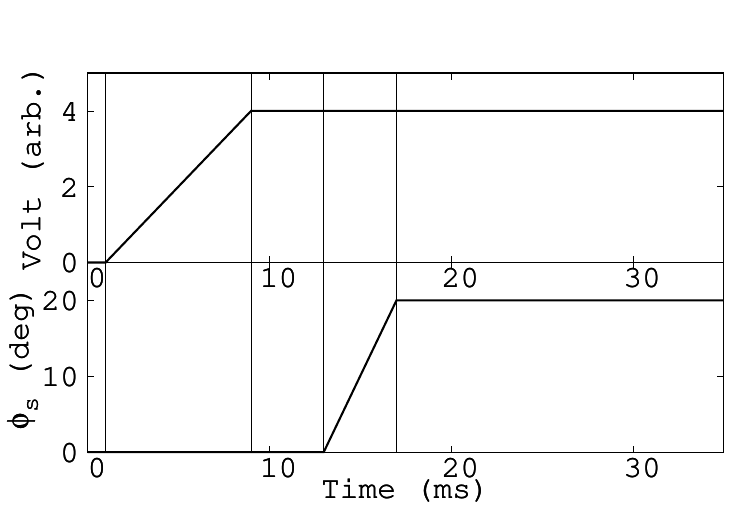}
}
\caption{\label{fig:awgfig1}AWG voltage and phase profile of two beams stacking.}
\end{figure*}

	RF gymnastics of the beam stacking experiment is controlled by the low level RF with an
Arbitrary Waveform Generator (AWG).
The output of the AWG controls amplitude and phase of the RF as a function of time.
Since it is more intuitive to specify amplitude and synchronous phase, $\phi_s$,
a script was prepared to convert these to AWG amplitude and phase outputs as a function of time.
Table~\ref{tab:awgtab1} and Fig.~\ref{fig:awgfig1} shows the input of the script for injection, 
acceleration, stacking and recapturing.

\begin{table}[h]
\caption{Specifications of RF gymnastics corresponding to Fig.~\ref{fig:awgfig1}. 
}
\label{tab:awgtab1}
\begin{ruledtabular}
\begin{tabular}{lcdr}
Process & duration\,[ms]\\
\colrule
\colrule
\multicolumn{2}{c}{Injection and acceleration of the first beam} \\
\colrule
constant AWG voltage of 4 (arb.)\\
\hspace{3mm} and $\phi_s$ of 20 (deg) & $t_\textrm{acc,1}$=6.55\\
linear ramp of $\phi_s$ to 0 (deg) & 4\\
stationary bucket for diagnostics & 1\\
linear ramp of AWG voltage to 1.18 (arb.) & 4\\
linear ramp of AWG voltage to 0.0 (arb.)  & 2\\
waiting for the second beam & 15.673\\
\colrule
\multicolumn{2}{c}{Injection and acceleration of the second beam} \\
\colrule
constant AWG voltage of 4 (arb.)\\
\hspace{3mm} and $\phi_s$ of 20 (deg) & $t_\textrm{acc,2}$\\
linear ramp of $\phi_s$ to 0 (deg)\\
\hspace{3mm} and AWG voltage to 1.18 (arb.) & 4\\
stationary bucket for diagnostics & 1\\
linear ramp of AWG voltage to 0.0 (arb.) & 2\\
coasting time & $t_\textrm{coast}$\\
\colrule
\multicolumn{2}{c}{Recapture and extract stacked beam} \\
\colrule
linear ramp of AWG voltage to 4 (arb.) & 8\\
stationary bucket for diagnostics & 4\\
linear ramp of $\phi_s$ to 20 (deg) & 4\\
acceleration toward aperture limit & 18\\
\end{tabular}
\end{ruledtabular}
\end{table}

Time duration for injection and acceleration of the second beam ($t_{acc,2}$ [ms] in 
Table~\ref{tab:awgtab1}) is taken as one of the main parameters in the experiment because it 
determines the final energy of the second beam $E_{acc,2}$, that is slightly different from the 
first beam to optimise the stacking process.
Note that the ramping pattern for the second beam is slightly different from the first one.
The larger $t_\textrm{acc,2}$ than $t_\textrm{acc,1}$ does not necessarily lead to the high energy of the second beam.

Table~\ref{tab:separationtab1} lists the final energy of the second beam as a function of $t_{acc,2}$ and the energy separation between the stacking energy (35.0206 MeV) and the second beam.

\begin{table}[h]
\caption{Energy separation of the stacking target energy (35.0206 MeV) and the second beam.}
\label{tab:separationtab1}
\begin{ruledtabular}
\begin{tabular}{lccdr}
$t_{acc,2}$ [ms] & $E_{acc,2}$ [MeV] & $\Delta E_{sep}$ [MeV]\\
\colrule
6.96 & 34.7852 & $-0.2354$\\
6.97 & 34.8204 & $-0.2006$\\
6.98 & 34.8556 & $-0.1650$\\
6.99 & 34.8909 & $-0.1297$\\
7.00 & 34.9262 & $-0.0944$\\
7.01 & 34.9751 & $-0.0591$\\
7.02 & 34.9968 & $-0.0238$\\
7.03 & 35.0321 & $+0.0115$\\
7.04 & 35.0674 & $+0.0468$\\
7.05 & 35.1028 & $+0.0822$\\
7.06 & 35.1382 & $+0.1176$\\
7.07 & 35.1735 & $+0.1529$\\
7.08 & 35.2089 & $+0.1883$\\
\end{tabular}
\end{ruledtabular}
\end{table}

Typically, there is a coasting time before recapturing ($t_{coast}$ [ms] in Table~\ref{tab:awgtab1}) of 303 ms for the Schottky measurement.
However, it is as short as 1 ms for some measurements, such as the measurement of the beam intensity right after the stacking.

The RF frequency of the recapturing process is chosen at the average of the first and second beams' RF frequency to ensure the maximum number of particles are captured.
This capturing process is necessary to make sure that no particle is circulating before the next set of accelerating cycles.
Since the main lattice magnet has a fixed field, particles can circulate in the machine for the beam lifetime.

The peak RF voltage at the cavity is approximately 4 kV when the AWG voltage is 4 (arb.).
This value slightly depends on the RF frequency, which is corrected by adding a frequency dependent weighting factor on the AWG output.

\begin{figure*}
\centering
\subfloat[Injection, acceleration and stacking of the one beam.]{
\includegraphics[width=\columnwidth]{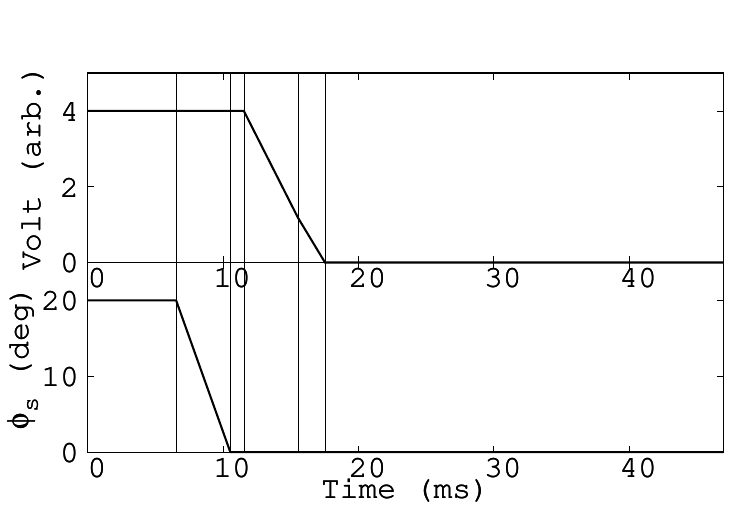}
}
\hspace{0mm}
\subfloat[Recapture of the one beam.]{
\includegraphics[width=\columnwidth]{figures/AWG/fig6r_main_eps_converted_to.pdf}
}
\caption{\label{fig:awgfig2}AWG voltage and phase profile of one beam.}
\end{figure*}

A separate RF pattern was prepared in order to measure the lifetime and momentum spread of a single beam.
The RF pattern parameters are described in Table~\ref{tab:awgtab2} and Fig.~\ref{fig:awgfig2}.

\begin{table}[h]
\caption{Specifications of RF gymnastics corresponding to Fig.~\ref{fig:awgfig2}
}
\label{tab:awgtab2}
\begin{ruledtabular}
\begin{tabular}{lcdr}
Process & duration\,[ms]\\
\colrule
\multicolumn{2}{c}{Injection and acceleration of the first beam} \\
\colrule
constant AWG voltage of 4 (arb.)\\
\hspace{3mm} and $\phi_s$ of 20 (deg) & 6.55\\
linear ramp of $\phi_s$ to 0 (deg) & 4\\
stationary bucket for diagnostics & 1\\
linear ramp of AWG voltage to 1.18 (arb.) & 4\\
linear ramp of AWG voltage to 0.0 (arb.)  & 2\\
waiting for the second beam & 15.673\\
coasting time & $t_\textrm{coast}$\\
\colrule
\multicolumn{2}{c}{Recapture and extract stacked beam} \\
\colrule
linear ramp of AWG voltage to 4 (arb.) & 8\\
stationary bucket for diagnostics & 4\\
linear ramp of $\phi_s$ to 20 (deg) & 4\\
acceleration toward aperture limit & 18\\

\end{tabular}
\end{ruledtabular}
\end{table}

The recapturing pattern is similar to the shown in Fig.~\ref{fig:awgfig1} (b).
However the RF frequency is chosen to match the revolution frequency of the single beam, rather than the average of the two beams.

	\subsection{\label{sec:expt_setup:schottky} Schottky Signals}
	Since stacked beams are initially coasting, obtaining their momentum
distribution, without potentially distorting excitation or extraction, requires
a measurement of the incoherent particle momenta. Since momentum and 
revolution frequency are connected with the slippage factor,
$\eta = \frac{1}{\gamma^2} - \frac{1}{\gamma_{t}^2}$, the relationship
\begin{align}
	\frac{\delta \omega}{\revangfreqnominal} = \slippage \frac{\delta p}{\momentumnominal},
\end{align}
can be used to infer the momentum distribution from a measurement of the
incoherent revolution frequency; $\revangfreqnominal$ is the nominal angular revolution
frequency, $\delta \omega$ is
the spread in revolution frequency, $\momentumnominal$ is the nominal momentum
and $\delta p$ is the momentum spread.
Incoherent revolution frequency distributions can be measured from the
spectra of longitudinal Schottky signals induced in capacitive pickups
\cite{CASschottky}. Such pickups typically have bandwidths
covering many revolution frequency harmonics, and are insensitive to the
coherent DC excitation of a coasting beam.

If the pickup is independent of the beams' transverse position, and its amplifier
chain has a flat frequency response, then the longitudinal Schottky spectra
is expected to resemble a series of broadening peaks centred on revolution
harmonics.
Any harmonic can be used to analyse the momentum distribution in principle, but
if spectra are obtained from the FFT of a time-sampled signal, then the broader
peaks provide a better momentum resolution up until adjacent peaks overlap. In
practice, the improved signal-to-noise-ratio (SNR) and the existence of external
noise can justify the analysis of lower frequency harmonics
\cite{nolden_instrumentation_2001}.

The power spectral density (PSD) in a small frequency band is proportional to
the number of protons with a revolution frequency contributing to that band.
Consequently, the integrated PSD of a single harmonic is proportional to the
total number of circulating protons. For a sampled signal, the PSD can be
estimated from the periodogram, which is the squared magnitude of the FFT
\cite{petre_stoica_spectral_2005}. To reduce the variance of each frequency
sample, it is possible to average periodograms. Where frequency resolution
can be sacrificed, a single time-series acquisition can be split into several
subsets and the PSD computed for each one. This is the basis of Bartlett's method, which computes the mean of the computed PSDs \cite{bartlett_periodogram_1950}.
The Welch method is a refinement, which allows adjacent subsets to overlap
and applies a window function to each one before computing the periodogram
\cite{welch_use_1967}; it
is used extensively in the following analysis.

	\subsection{\label{sec:expt_setup:diagnostics} FAB Bunch Monitor}
	\label{method:bunchmonitor}

Longitudinal beam profiles for Schottky analysis were acquired using the 
FAB \cite {fab_ipac}.
The FAB is a capacitive pickup with a wide rectangular aperture, measuring $\SI{1000}{mm}$ 
horizontally and 
$\SI{75}{mm}$ vertically.
This covers the beam orbit from injection to extraction, and its output voltage is
independent of the beam's transverse position.  

The monitor was terminated using a SA220-F5 low-noise amplifier ((C) NF Corporation) with a $\SI{1}{\Mohm}$ input impedance.  
The power gain of the amplifier is $\SI{46}{dB}$ and the upper bandwidth is $\SI{80}{\MHz}$. 
The capacitance of the monitor is about $\SI{270}{pF}$ including the amplifier and vacuum feed-through.  

	\subsection{\label{sec:expt_setup:daq} Data Acquisition}
	In order to use the Schottky signals to measure the momentum spread of the coasting beam, it is critical that the noise in the measurement is reduced as much as possible, or the Schottky signal will be obscured. 
In preliminary measurements, it was not possible to identify a clear signal as there was too much 
noise to make out a Schottky peak. Much of this noise originated from the long
\SI{20}{}- \SI{30}{\m} cable between the FAB and the control room. Moving the oscilloscope to the 
machine room, and connecting it via a $\sim\SI{4}{\m}$ cable, reduced the noise 
sufficiently to see a Schottky signal from the coasting beam. To observe as high a harmonic 
Schottky peak as possible, the noise in the frequency domain was further reduced by using a low-pass filter and increasing the sample rate.

\section{\label{sec:meas} Measurement}

	\subsection{\label{sec:meas:lifetime} Beam Lifetime}
	An important outcome of the experiments was to determine the efficiency of beam stacking in capturing both beams. In order to understand the losses, it was necessary to measure the beam lifetime. Two complementary measurement techniques were used to measure the beam lifetime: one based on the decay of the bunched beam intensity and the second based on the Schottky signal of a coasting beam.

\begin{figure}
\centering
\includegraphics[width=\columnwidth]{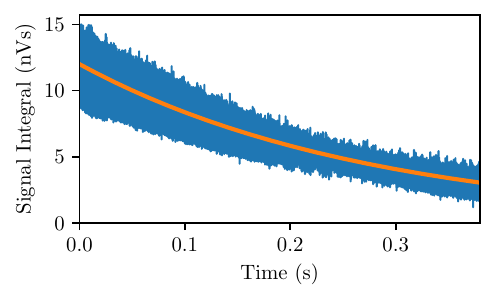}% Here is how to import EPS art
\caption{\label{fig:lifetime}Beam lifetime measurement with the beam stored in a stationary bucket. The time constant of the exponential fit (solid orange line) provides a measure of lifetime.}
\end{figure}

\subsubsection*{Bunched beam Lifetime}

A measurement of beam lifetime was made by storing the bunch in a stationary bucket at the stacked beam energy. The FAB monitor signal was integrated turn-by-turn. 
In order to measure the effect of noise, the same measurement was made without injecting beam. The bunch intensity is given by the integrated, turn-by-turn FAB monitor signal, with the mean noise subtracted. A fit was made to a function of the form $N \textrm{exp}(-t/\tau)$. From the 10 sets of data obtained, the mean lifetime was found to be $\tau = (270 \pm 11)$\,ms where the uncertainty is the standard deviation of the measurements. An example measurement is shown in Fig.\,\ref{fig:lifetime}. Another approach to calculate the bunched beam lifetime, based on the magnitude of the fundamental Fourier harmonic, is described in Appendix~\ref{apdx:lifetime}.

\subsubsection*{Coasting Beam Lifetime}

\begin{figure}
	\centering
	\includegraphics{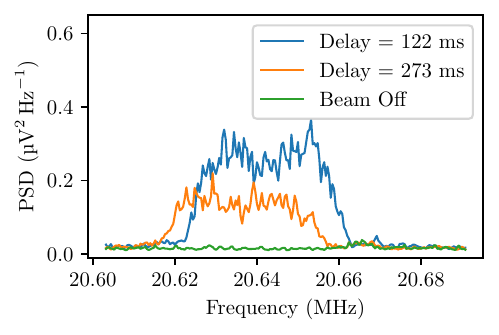}
	\caption{PSD estimated by the Welch method for a single injected bunch,
		with two delay settings. Each beam-on acquisition is the mean of five
		measurements, and the beam-off is a mean of two.}
	\label{fig:lifetime:schottky_01}
\end{figure}

The beam lifetime was also estimated from the Schottky signals after coasting
for different periods of time. The integrated PSD over a single Schottky harmonic
is proportional to the total number of circulating protons, $\numprotons(t)$.
The lifetime can be estimated by assuming an exponential decay with
decay constant $\tau$, and computing the integrated PSD at two times.
Coasting beam data was acquired with the FAB for
\SI{50}{\ms} with a delay setting of \SI{122}{\ms} and \SI{272}{\ms};
starting from approximately \SI{80}{\ms} and \SI{230}{\ms} after the RF amplitude reached
\SI{0}{\volt}. Figure~\ref{fig:lifetime:schottky_01} shows the PSD computed 
from the entire acquisition window with the Welch method and 20 chunks. From this,
it is clear that, in addition to the anticipated reduction in amplitude, the
revolution frequency has also reduced; this effect has been attributed to
small-angle scattering from residual gas. 

\begin{figure}
	\centering
	\includegraphics{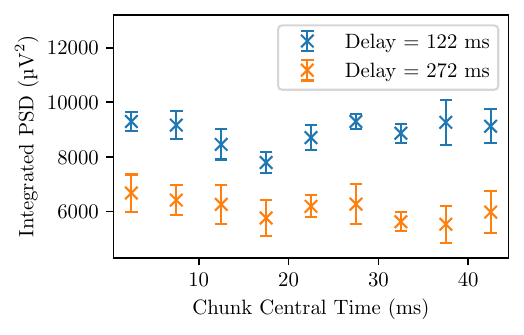}
	\caption{Integrated PSD of a single Schottky harmonic computed with different
		delay settings, from \SI{5}{\ms} chunks
		of a \SI{50}{\ms} acquisition window. Points are a mean of five beam-on acquisitions, 
		with the mean of two beam-off acquisitions subtracted; uncertainties are the
		standard error in the beam-on mean. The dashed box indicates a pair of chunks
		used to compute a lifetime.}
	\label{fig:lifetime:schottky_02}
\end{figure}

The acquisition duration of \SI{50}{\ms} was not expected to be a
negligible fraction of the beam lifetime. To avoid systematic distortions
from this large acquisition window, each dataset was
split into chunks of \SI{5}{\ms} and the integrated PSD computed for each chunk.
This duration was expected to be negligible compared to the beam lifetime.
Since both datasets had the same duration, pairs of corresponding chunks with
a constant time-separation were identified and used to compute a lifetime; see 
Figure~\ref{fig:lifetime:schottky_02}. A weighted mean was obtained from the
computed lifetimes on several days, which ranged from $\lifetime=\SI{300(10)}{\ms}$
to $\SI{390(30)}{\ms}$.

Whilst the lifetimes of bunched and unbunched beams differed, and the latter varied by up to
\SI{30}{\percent} day-to-day, all measurements were consistently above \SI{270}{\ms}.
Crucially, this meant that beam losses of $\leq \SI{12}{\percent}$ are expected for the first beam,
because of the extra time ($\sim \SI{33}{\ms}$) it spends circulating the machine. More than 
\SI{25}{\percent} of the first beam is expected to survive for the overall $\sim \SI{350}{\ms}$ 
spent in the ring, allowing faster loss mechanisms associated with the stacking process to be 
identified.

	\subsection{\label{sec:beam stacking} Stacked Beam Momentum Spread}

\begin{figure}[h]
	\centering
	\subfloat[][]{
	\includegraphics[width=0.48\linewidth]{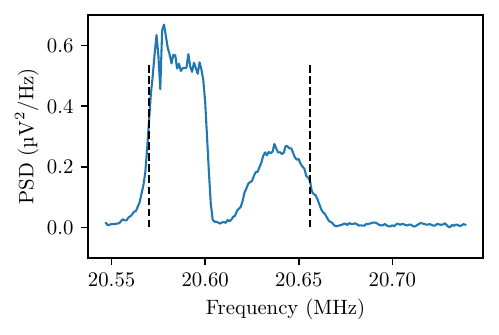}\hfill
	\label{fig:meas:momentum:psd:696}
	}
	\subfloat[][]{
	\includegraphics[width=0.48\linewidth]{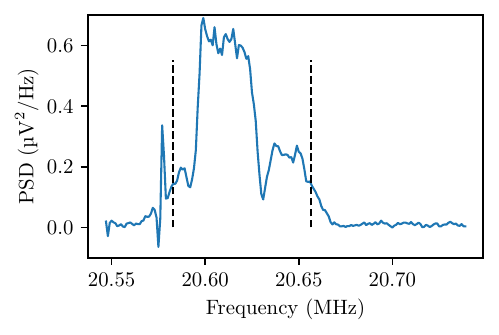}
	\label{fig:meas:momentum:psd:699}
	}\\
	\subfloat[][]{
	\includegraphics[width=0.48\linewidth]{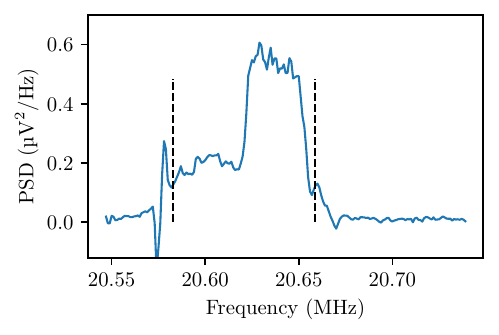}\hfill
	\label{fig:meas:momentum:psd:702}
}
	\subfloat[][]{
	\includegraphics[width=0.48\linewidth]{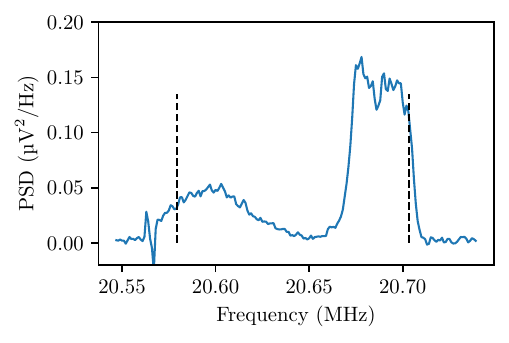}
	\label{fig:meas:momentum:psd:708}
}
\caption{PSD of the eighth revolution harmonic, estimated with
		 Welch's method for final energy separations of 
		 a) \SI{-235}{\keV}, b) \SI{-130}{\keV}, c) \SI{-24}{\keV}
		 and d) \SI{+180}{\keV}. Dashed lines indicate
		 \SI{90}{\%} widths.}
\label{fig:meas:momentum:psd}
\end{figure}

Beam stacking was optimised by holding the final energy of the first beam constant,
and varying the final energy of the second beam by modifying the RF settings for the second acceleration cycle.
Five beam-on and two beam-off measurements were acquired for each final energy. 
The PSD was computed for each dataset using the Welch method, and the mean
beam-off PSD subtracted from the beam-on. Figure~\ref{fig:meas:momentum:psd}
shows the resulting PSD for several acceleration times of the second beam.  As seen in the figure, especially Fig.~\ref{fig:meas:momentum:psd} (a) and (d), two peaks appear that can be identified as the two beams. This indicates successful injection and stacking in the FFA machine.

The two beams shown in Fig.~\ref{fig:meas:momentum:psd} are not equally 
populated and have different momentum distributions. This makes estimating
the overall momentum spread challenging, as all width estimators will be
distribution dependent. For example, computing the 90\% width will
systematically measure smaller widths when the less populous beam surrounds
the other, as  in Fig.~\ref{fig:meas:momentum:psd:699}. To investigate
the impact of this, and ensure that it does not affect conclusions,
an additional method is described in Appendix~\ref{apdx:width:lines}.
Both show a similar distribution of width with energy gap.

\begin{figure}[h]
	\centering
	\includegraphics{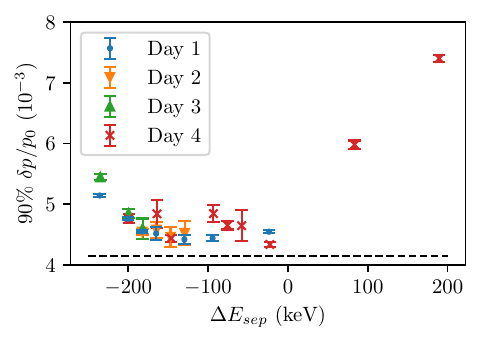}
	\caption{90\% width computed from the PSD of the eighth revolution harmonic.
	Error bars are the standard error on the mean of five datasets. The dashed line
	indicates twice the mean 90\% width of a single-beam.
	}
	\label{fig:meas:momentum:width}
\end{figure}

90\% widths were computed from a cumulative trapezoid integral of PSD around the
eighth harmonic. Linear interpolation was used to identify
frequencies delimiting 5\% and 95\% of the  integral.
Figure~\ref{fig:meas:momentum:psd} shows some indicative
width markers, and Fig.~\ref{fig:meas:momentum:width} shows the mean
90\% width of five beam-on datasets. Uncertainties were estimated from
the standard error on the mean, and repeat measurements were made over
a number of days to highlight systematic variations from environmental
or machine conditions. Whilst variations are observed, 
minima are consistently found with an energy separation of \SI{-180}{\keV}.

Since stacking cannot reduce longitudinal emittance, the stacked
beam must have a momentum spread equal to or greater than the sum of
its component beams. 
The optimum $\delta p / p_0$ was obtained with an energy gap close to
\SI{-180}{\keV}, however Fig.~\ref{fig:meas:momentum:width} demonstrates
that this exceeded twice the single-beam momentum-spread. This effective
emittance growth is attributed to processes described in
Section~\ref{sec:simulation}.

\subsection{Stacked Beam Intensity}

Although a similar RF pattern was used for both beams, the amplitude
of the first beam in Fig.~\ref{fig:meas:momentum:psd}(a) is significantly reduced
compared to that of the second beam. Whilst the RF knock-out effect was expected to
cause beam-loss, the extent of this was not known in advance.
Provided the two peaks are sufficiently separated, the fractional loss can
be estimated from the ratio between the areas under the individual Schottky PSD
harmonics.

These were estimated using the distributions shown in
Figs.~\ref{fig:meas:momentum:psd}(a)~and~\ref{fig:meas:momentum:psd}(d).
The ratios between second and first beams are \SI{0.51(0.04)}{} and
\SI{0.47(0.05)}{}, which are both consistent with
half of the first beam being lost as the second was accelerated.
Since the measured lifetime predicts $\leq \SI{12.1}{\%}$ losses over
the \SI{33}{\ms} coasting time, a separate loss mechanism, such as RF knock-out,
was required to explain this observation.

\section{\label{sec:discussion} Discussion}
\label{discussion}

The Schottky signal shows that the momentum spread of the stored beam is similar for the first and the second beams for the whole process.
The beam intensity of the first beam is, however, significantly reduced (about a half) when the second beam is accelerated.
In order to identify the source of the beam intensity reduction, we compared the beam intensity of 
the first and second beams in the 4 different beam-filling patterns described in
Table~\ref{tab:cases1to4}.
\begin{table}[b]
	\centering
	\begin{ruledtabular}
		\caption{Four different beam-filling patterns.}
		\label{tab:cases1to4}
	\begin{tabular}{l|cc}
			   & First Beam RF Bucket & Second Beam RF Bucket \\ \colrule
		Case 1 & Filled           & None             \\
		Case 2 & Filled           & Empty            \\
		Case 3 & Empty            & Filled           \\
		Case 4 & Filled           & Filled          
	\end{tabular}
	\end{ruledtabular}
\end{table}

The first pattern (named Case 1) was a reference.
The beam was captured in the RF bucket and accelerated to the top energy.
No RF or beam was filled for the second beam.
The rest of the three cases have an identical RF pattern for the both beams, but the beam
filling was different.
Case 2 had the first beam injected into the RF bucket, whilst the second RF bucket was
kept empty.
Case 3 started with an empty RF bucket with the same RF pattern for acceleration. A beam was
then injected into the second RF bucket when the first empty bucket reached the top energy.
In Case 4, both RF buckets were filled, as for the nominal beam stacking case.
\begin{figure*}[htb]
\centering
\subfloat[Case 1]{
	\includegraphics[width=0.48\columnwidth]{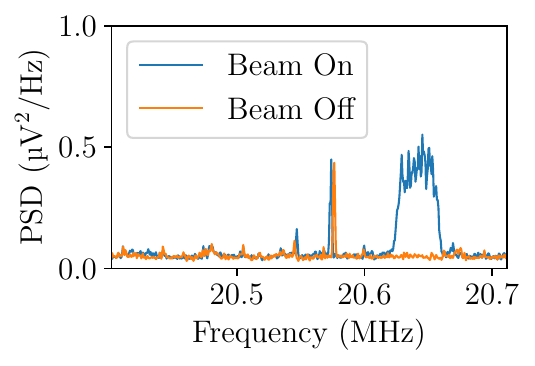}
}
\subfloat[Case 2]{
	\includegraphics[width=0.48\columnwidth]{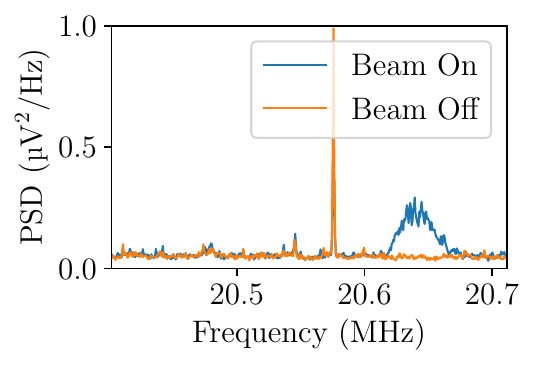}
}
\hspace{0mm}
\subfloat[Case 3]{
	\includegraphics[width=0.48\columnwidth]{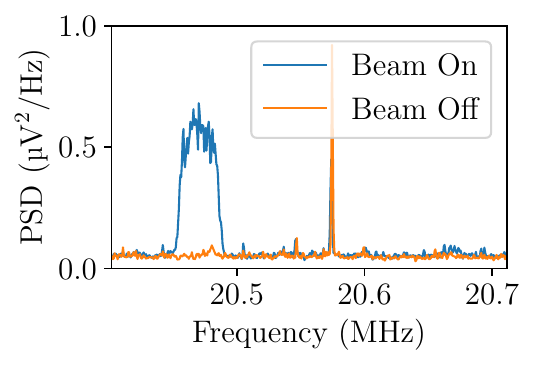}
}
\subfloat[Case 4]{
	\includegraphics[width=0.48\columnwidth]{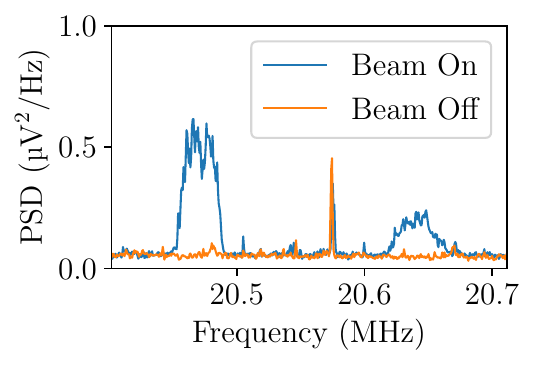}
}
\caption{Schottky signals with the four filling patterns listed in
	Table~\ref{tab:cases1to4}.}
\label{fig:cases1to4_schottky}
\end{figure*}

The Schottky signals of all four cases are shown in Fig.~\ref{fig:cases1to4_schottky}.
Cases 1 and 3 are similar, because they differed only in when the Schottky signal was observed.
The sweep of an empty bucket before the acceleration of the second beam did not affect the second 
beam.
In Cases 2 and 4, on the other hand, the intensity of the first beam was reduced, 
whether the second RF bucket was filled or not.
RF knock-out \cite{terwilliger_radio_1956, jones_experiments_1957} is one possible source of the 
intensity reduction of the first beam. 

\subsection{RF knock-out}
In the usual operation of accelerators, the RF frequency $\omega_\textrm{RF}$ for acceleration or storage has to be synchronised with the revolution frequency of the particles $\omega_\textrm{rev}$ or its harmonics $h\omega_\textrm{rev}$.
During the process of beam stacking, that is not necessarily the case.
The RF bucket has a frequency synchronised with the speed of the accelerating beam.
The previously accelerated coasting beam circulates at the top energy, and always has a higher revolution
frequency than the RF frequency divided by the harmonic number $h$.
Here we assume that beam stacking is performed below the transition energy.

Particles in the coasting beam see the voltage at the RF cavity as
\begin{equation}
\begin{split}
V_\textrm{gap} & = V_0 \cos\left( \omega_{RF}nT_\textrm{rev} \right) \\
& = V_0 \cos \left( 2\pi \frac{\omega_\textrm{RF}}{\omega_\textrm{rev}} n\right) \\
& = V_0 \cos \left( 2\pi \frac{\Delta\omega}{\omega_\textrm{rev}} n\right),
\end{split}
\end{equation}
where 
\begin{equation*}
\Delta\omega = h\omega_\textrm{rev}-\omega_\textrm{RF},
\end{equation*}
and $n$ is an integer number, $V_0$ is the amplitude of the RF voltage, and $T_\textrm{rev}$ is 
the revolution time of the coasting beam.
Sudden energy gain at the RF cavity shifts the equilibrium orbit proportional to the dispersion function $D_x$ at the location of the cavity, that is
\begin{equation}
\begin{split}
\Delta x & = -D_x \frac{\Delta p}{p_0} \\
& =-\frac{\gamma D_x}{\left( 1+\gamma \right)}\frac{\Delta E_k}{E_{k0}} \\
& =-\frac{\gamma D_x V_0 }{\left( 1+\gamma \right) E_{k0}}
\cos \left( 2\pi \frac{\Delta\omega}{\omega_\textrm{rev}} n\right),
\end{split}
\end{equation}
where $\gamma$ is the Lorentz factor, $E_{k0}$ is the nominal kinetic energy and $\Delta E_k$ is its deviation.

When the frequency of the displacement is synchronised with the frequency of horizontal betatron oscillation $\omega_{h\beta}$,
\begin{equation}
    \frac{\Delta\omega}{\omega_\textrm{rev}} = \frac{h\omega_\textrm{rev}-\omega_{RF}}{\omega_\textrm{rev}}=\frac{\omega_{h\beta}}{\omega_\textrm{rev}}-Q_I \text{ or } Q_I-\frac{\omega_{h\beta}}{\omega_\textrm{rev}}
\end{equation}
where $Q_I$ is the nearest integer of $\frac{\omega_{h\beta}}{\omega_\textrm{rev}}$, a resonance occurs, and the displacement will be coherently accumulated.
Eventually, a particle is lost when its betatron amplitude exceeds either the physical or dynamic 
aperture.

Unlike the synchro-betatron resonance, the effects are in the transverse phase space. 
In longitudinal phase space, the coasting beam is unaffected by the accelerating RF bucket
until $\omega_\textrm{RF}$ approaches $\omega_\textrm{rev}$.

\begin{figure*}[htb]
\centering
\subfloat[Sweep of the RF frequency in blue line and change of horizontal tune in orange curve. Small oscillations of the tune is due to a numerical error.]{
\includegraphics[width=\columnwidth]{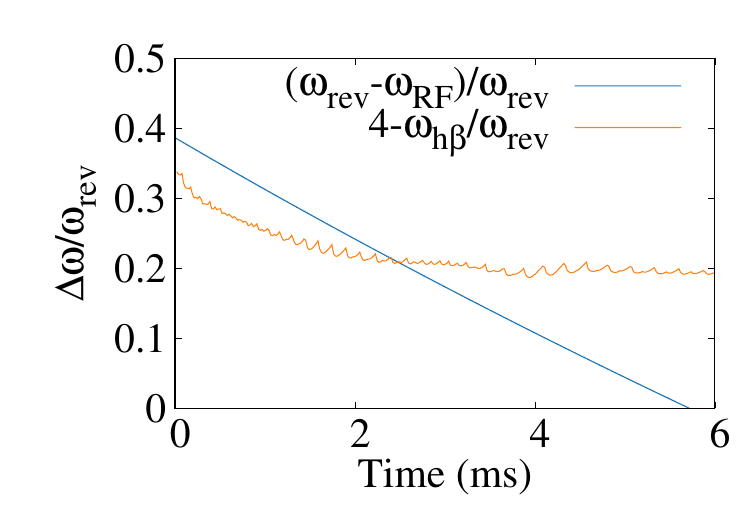}% Here is how to import EPS art
}
\hspace{0mm}
\subfloat[Horizontal emittance evolution. When the RF frequency is equal to horizontal tune, the emittance growth is observed.]{
\includegraphics[width=\columnwidth]{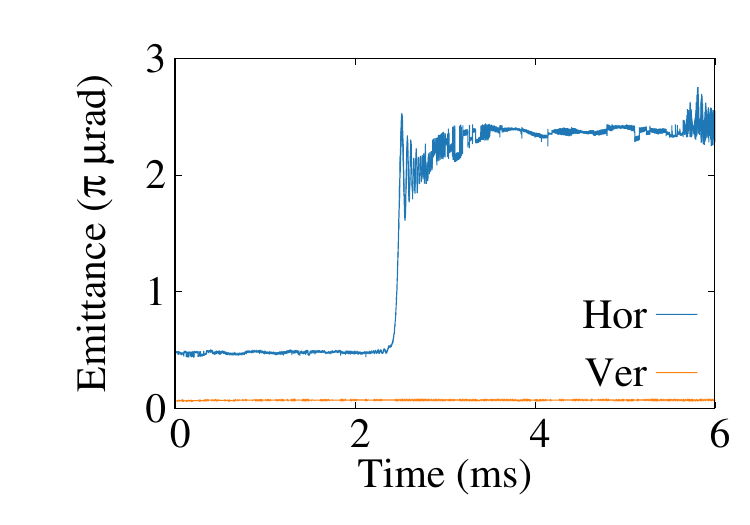}% Here is how to import EPS art
}=\caption{\label{fig:knockout}Simulation results of the RF knock-out in the KURNS FFA.}
\end{figure*}

\subsection{Simulation}

A 6D phase space simulation was carried out to see if the beam intensity reduction can be reproduced through simulation of the RF knock-out process.
We used the 3D field map of the KURNS FFA triplet magnet, which was created in the OPERA 3D software.
The ring is made of 12 identical field maps and the symmetry is preserved.
The magnet model is accurate to the real machine, which means the scaling principle is not fully satisfied.
The total transverse tune varies for around 0.1. 
For the validation of the RF knock-out, the correct modelling of the longitudinal motion is essential.
A moving RF bucket is created by increasing the RF frequency and integrating the RF phase at the cavity.

Figure \ref{fig:knockout} shows that emittance growth occurs when the resonance condition is satisfied at around 2.5 ms.
Although the horizontal physical aperture in an FFA is huge, the limited dynamic aperture due to intrinsic nonlinearities of the magnets causes the beam loss when the sizable emittance growth occurs.
After the particles go through those resonances, no other effects in the horizontal direction are seen.
As expected, there is no effect in the vertical phase space.

\subsection{Mitigation}
Normally, the optics of an FFA has no dispersion-free section.
The knock-out effects are inevitable during beam stacking in an FFA.
Several mitigation measures were discussed previously \cite{kerst_suppression_1957}.
One proposal was to reduce or switch off the RF voltage at the time when the resonance condition is satisfied.
Although the RF knock-out effects could be mitigated, a mismatch in the longitudinal phase space is unavoidable.
After the reduction of the RF voltage, dipole oscillations of the bunch could be a source of beam loss.

The most promising mitigation method is to use cancellation of the horizontal amplitude growth across several RF cavities.
A simple example is to place the RF cavities at locations in the ring which satisfy the condition that the betatron oscillation phase difference between cavity positions in the horizontal direction is $\pi$, and magnitude of the dispersion function is the same at each RF cavity location.
The RF frequency should be the same and the phase difference should be synchronised with the speed of the coasting energy particles, not the accelerating beam.

It is not always easy to find two locations for the RF cavities which satisfy these conditions.
When there are more than 2 RF cavities in the ring, cancellation of the knock-out effects can be achieved if the RF voltage of each RF cavity can be adjusted.
For example, let us assume that three RF cavities are located with the same dispersion function and the phase differences between the first and the second and the second and the third are equal.
In that case, cancellation happens when the voltage satisfies the following conditions.
\begin{equation}
\begin{split}
    V_3 &=V_1 \\
    V_2 &=2 V_1\cos \Delta \phi,
\end{split}
\end{equation}
where $V_i$ is the voltage of the cavity with the index $i$ and $\Delta \phi$ is the horizontal phase advance between cavities.

When the beam intensity is high, space charge tune spread is not negligible.
Phase advance between cavities will not be a single value.
Further study on the mitigation measure of the RF knock-out is under consideration.

\section{\label{sec:conclusion} Conclusion}
Beam stacking is one way to increase the average and peak intensity (or the number of particles per bunch) of hadron beams without enhancing space charge effects at the injection energy.
In addition, it separates the accelerator cycle and the user cycle, which is beneficial for applications such as proton drivers of spallation neutron facilities. In such facilities, the neutron users prefer low repetition rates such as 10 to 25 Hz, but it is easier for the beam loss mitigation purposes in the accelerators to operate a high repetition rate such as 100 to 200 Hz.

Since the beam stacking we consider here uses stacking in the longitudinal phase space, in the momentum direction, a wide momentum acceptance is necessary in the accelerator.
An FFA accelerates the beam from the injection to extraction energy with fixed field magnets.
The momentum acceptance is given by the corresponding momentum range and is typically huge and time-independent.
Beam stacking at the extraction energy, as well as acceleration of consecutive beams, can be carried out in a single machine using an FFA.

It is inevitable to have a larger longitudinal emittance and momentum spread, which linearly increases with the number of stacked beams.
By simulation, additional sources of the emittance increase due to non-adiabatic processes such as scattering are clarified.
Detailed simulations were performed to minimize the final momentum spread, which specifies the RF voltage and phase as a function of time.
This optimisation is critical, especially at the end of acceleration of consecutive beams.

We have demonstrated experimentally the beam stacking of two beams at KURNS.
The Schottky scan signal was used to measure the momentum spread and the intensity of the coasting beam after the stacking process.
The results show that beam stacking can be done with only a slight increase of the momentum spread in Fig.
\ref{fig:meas:momentum:width}.

The beam intensity of the first beam was, however, reduced by a factor 0.51${\pm}$0.04 at the setting having the lowest momentum spread.
Beam lifetime measurements indicate an expected loss of a factor 0.88.
Therefore this rate of loss cannot be explained by the beam lifetime.

We suspect that the RF knock-out is the source of the beam loss.
6D simulation has been done to see the effects.
Mitigation measures such as the phasing of multiple RF cavities are discussed.
Further experiment to investigate the proper measure of mitigation is summarised.

\appendix

\section{Another way to estimate a width in the PSD}\label{apdx:width:lines}
Schottky signal does not always clearly show a boundary where the beam signal exists.
It becomes even harder to identify when two beams get close and some part of the first beam is phase displaced.
In Section~\ref{sec:beam stacking}, we took the measure of 90\% width to define the frequency spread.
In this Appendix, we will discuss another way to define the width of the PSD and the momentum spread.
We called this method a straight-line fitting because of the procedure we will explain below.

\begin{figure*}[htb]
    \centering
    \subfloat[]{\includegraphics{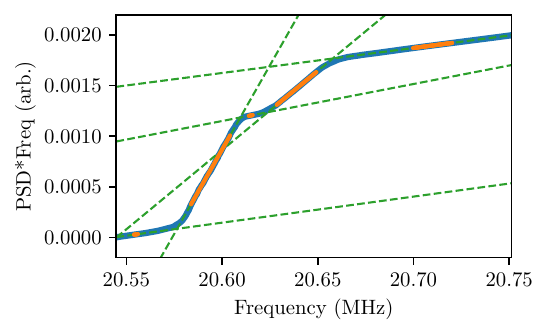}}
    \subfloat[]{\includegraphics{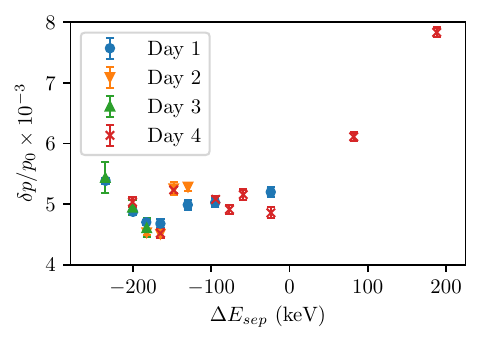}}
    \caption{(a) Example of straight-line fit for noise and beam. The orange areas are used to fit a straight line. Intersections of dashed lines correspond to the frequency boundary of the beams. (b) Momentum spread calculation 
    by straight-line fitting. Error bars are the standard error on the mean of five datasets.
    	%\textcolor{red}{$p \rightarrow p_0$, e-3 to e+3 or similar}
    	}
    \label{fig:appendix_f15}
\end{figure*}

After obtaining the PSD for each measurement, the signal is integrated over the frequency axis.
Assuming the noise level is independent of frequency,
the integrated value increase linearly over the frequency axis as shown in Fig.~\ref{fig:appendix_f15} (a).
When the distribution of the momentum is uniform and the Schottky signal has a rectangular shape, at the frequency where the Schottky signal starts appearing, the integrated value will linearly be rising with a steeper slope until the upper edge of the frequency.
The integrated value becomes flatter again where only the noise signal contributes to the integration.
%When there is a finite noise level, this makes the integrated value linearly increase also but with a shallower gradient.

By fitting several straight lines for noise and beam and finding the intersection points, the beam width can be calculated.
Figure~\ref{fig:appendix_f15} (b) shows the consistent result with Fig.~\ref{fig:meas:momentum:width}.
Assignment of the straight-ling fitting was done by eye for each data.

\section{Lifetime from Bunch Spectrum}\label{apdx:lifetime}
\begin{figure*}[htb]
    \centering
    \subfloat[]{\includegraphics{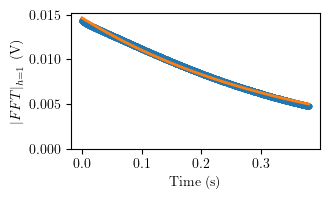}}
    \subfloat[]{\includegraphics{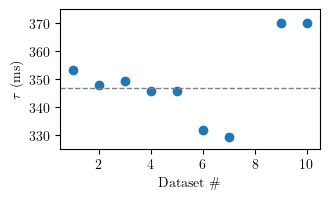}}
    \caption{(a) Evolution of the fundamental harmonic of the beam spectrum (blue) and exponential fit (orange). (b) Lifetime, $\tau$, found for each dataset. The dashed line is the mean value. Note, the example shown on the left corresponds to the first dataset
    	}
    \label{fig:appendix_f16}
\end{figure*}

The bunched beam lifetime reported in Sec.\,\ref{sec:meas:lifetime} is sensitive to noise in the FAB monitor signal. Another measure of lifetime can be made based on the magnitude of the fundamental harmonic of the bunch spectrum. It is observed that when the RF is on, but no beam is injected, there is no noise component in the FAB monitor data at the fundamental frequency. It follows, assuming there is no significant change in the shape of the distribution while the beam is stored, that the magnitude of the fundamental harmonic is proportional to the bunched beam intensity. 

A spectrogram of the data was calculated in order to obtain the magnitude at the fundamental frequency as a function of time. A fit of the resulting data was made to a function of the form $N \textrm{exp}(-t/\tau)$. An example of an individual measurement is shown in Fig.\,\ref{fig:appendix_f16}(a). The mean result for the complete set of measurements, shown in Fig.\,\ref{fig:appendix_f16}(b), is $\tau = (347 \pm 15)$\,ms where the uncertainty is the standard deviation of the measurements. It can be concluded that the lifetime is too long to explain the observed beam loss during the beam stacking process.

\begin{acknowledgments}
The work described here was made possible by grants from UK Research and Innovation.
The authors express their sincere gratitude to all the KURNS staff for their great cooperation.
\end{acknowledgments}

\bibliography{apssamp}

\providecommand{\noopsort}[1]{}\providecommand{\singleletter}[1]{#1}%
\begin{thebibliography}{30}%
\makeatletter
\providecommand \@ifxundefined [1]{%
 \@ifx{#1\undefined}
}%
\providecommand \@ifnum [1]{%
 \ifnum #1\expandafter \@firstoftwo
 \else \expandafter \@secondoftwo
 \fi
}%
\providecommand \@ifx [1]{%
 \ifx #1\expandafter \@firstoftwo
 \else \expandafter \@secondoftwo
 \fi
}%
\providecommand \natexlab [1]{#1}%
\providecommand \enquote  [1]{``#1''}%
\providecommand \bibnamefont  [1]{#1}%
\providecommand \bibfnamefont [1]{#1}%
\providecommand \citenamefont [1]{#1}%
\providecommand \href@noop [0]{\@secondoftwo}%
\providecommand \href [0]{\begingroup \@sanitize@url \@href}%
\providecommand \@href[1]{\@@startlink{#1}\@@href}%
\providecommand \@@href[1]{\endgroup#1\@@endlink}%
\providecommand \@sanitize@url [0]{\catcode `\\12\catcode `\$12\catcode
  `\&12\catcode `\#12\catcode `\^12\catcode `\_12\catcode `\%12\relax}%
\providecommand \@@startlink[1]{}%
\providecommand \@@endlink[0]{}%
\providecommand \url  [0]{\begingroup\@sanitize@url \@url }%
\providecommand \@url [1]{\endgroup\@href {#1}{\urlprefix }}%
\providecommand \urlprefix  [0]{URL }%
\providecommand \Eprint [0]{\href }%
\providecommand \doibase [0]{https://doi.org/}%
\providecommand \selectlanguage [0]{\@gobble}%
\providecommand \bibinfo  [0]{\@secondoftwo}%
\providecommand \bibfield  [0]{\@secondoftwo}%
\providecommand \translation [1]{[#1]}%
\providecommand \BibitemOpen [0]{}%
\providecommand \bibitemStop [0]{}%
\providecommand \bibitemNoStop [0]{.\EOS\space}%
\providecommand \EOS [0]{\spacefactor3000\relax}%
\providecommand \BibitemShut  [1]{\csname bibitem#1\endcsname}%
\let\auto@bib@innerbib\@empty
%</preamble>
\bibitem [{\citenamefont {Hotchi}(2021)}]{AAPPSBul}%
  \BibitemOpen
  \bibfield  {author} {\bibinfo {author} {\bibfnamefont {H.}~\bibnamefont
  {Hotchi}},\ }\bibfield  {title} {\bibinfo {title} {High-power proton
  accelerators for pulsed spallation neutron sources},\ }\href
  {https://doi.org/10.1007/s43673-021-00025-0} {\bibfield  {journal} {\bibinfo
  {journal} {AAPPS Bull}\ }\textbf {\bibinfo {volume} {31}},\ \bibinfo {pages}
  {23} (\bibinfo {year} {2021})}\BibitemShut {NoStop}%
\bibitem [{\citenamefont {Henderson}\ and\ \citenamefont {et.
  al.}(2014)}]{SNS}%
  \BibitemOpen
  \bibfield  {author} {\bibinfo {author} {\bibfnamefont {S.}~\bibnamefont
  {Henderson}}\ and\ \bibinfo {author} {\bibnamefont {et. al.}},\ }\bibfield
  {title} {\bibinfo {title} {The spallation neutron source accelerator system
  design},\ }\href@noop {} {\bibfield  {journal} {\bibinfo  {journal} {Nucl.
  Instr. Meth. Phys. Res. Section A}\ }\textbf {\bibinfo {volume} {763}},\
  \bibinfo {pages} {610} (\bibinfo {year} {2014})}\BibitemShut {NoStop}%
\bibitem [{\citenamefont {Hotchi}\ and\ \citenamefont {et. al.}(2009)}]{JPARC}%
  \BibitemOpen
  \bibfield  {author} {\bibinfo {author} {\bibfnamefont {H.}~\bibnamefont
  {Hotchi}}\ and\ \bibinfo {author} {\bibnamefont {et. al.}},\ }\bibfield
  {title} {\bibinfo {title} {Beam commissioning of the 3-gev rapid cycling
  synchrotron of the japan proton accelerator research complex},\ }\href@noop
  {} {\bibfield  {journal} {\bibinfo  {journal} {Phys. Rev. ST Accel. Beams}\
  }\textbf {\bibinfo {volume} {12}},\ \bibinfo {pages} {040402} (\bibinfo
  {year} {2009})}\BibitemShut {NoStop}%
\bibitem [{muo(2024)}]{muoncollider}%
  \BibitemOpen
  \href@noop {} {\bibinfo {title} {\text{IMCC} and \text{MuCol} annual meeting
  2024}} (\bibinfo {year} {2024}),\ \bibinfo {note}
  {\url{https://indico.cern.ch/event/1325963}}\BibitemShut {NoStop}%
\bibitem [{\citenamefont {Lagrange}\ \emph {et~al.}(2019)\citenamefont
  {Lagrange}, \citenamefont {Adams}, \citenamefont {Cavanagh}, \citenamefont
  {Gardner}, \citenamefont {Griffin-Hicks}, \citenamefont {Jones},
  \citenamefont {Kelliher}, \citenamefont {Letchford}, \citenamefont {Machida},
  \citenamefont {Pine}, \citenamefont {Prior}, \citenamefont {Rogers},
  \citenamefont {Thomason}, \citenamefont {Warsop}, \citenamefont {Williamson},
  \citenamefont {Pasternak}, \citenamefont {Pozimski}, \citenamefont {Brown},\
  and\ \citenamefont {Rees}}]{isisii}%
  \BibitemOpen
  \bibfield  {author} {\bibinfo {author} {\bibfnamefont {J.-B.}\ \bibnamefont
  {Lagrange}}, \bibinfo {author} {\bibfnamefont {D.~J.}\ \bibnamefont {Adams}},
  \bibinfo {author} {\bibfnamefont {H.~V.}\ \bibnamefont {Cavanagh}}, \bibinfo
  {author} {\bibfnamefont {I.~S.~K.}\ \bibnamefont {Gardner}}, \bibinfo
  {author} {\bibfnamefont {P.~T.}\ \bibnamefont {Griffin-Hicks}}, \bibinfo
  {author} {\bibfnamefont {B.}~\bibnamefont {Jones}}, \bibinfo {author}
  {\bibfnamefont {D.~J.}\ \bibnamefont {Kelliher}}, \bibinfo {author}
  {\bibfnamefont {A.~P.}\ \bibnamefont {Letchford}}, \bibinfo {author}
  {\bibfnamefont {S.}~\bibnamefont {Machida}}, \bibinfo {author} {\bibfnamefont
  {B.~G.}\ \bibnamefont {Pine}}, \bibinfo {author} {\bibfnamefont {C.~R.}\
  \bibnamefont {Prior}}, \bibinfo {author} {\bibfnamefont {C.~T.}\ \bibnamefont
  {Rogers}}, \bibinfo {author} {\bibfnamefont {J.~W.~G.}\ \bibnamefont
  {Thomason}}, \bibinfo {author} {\bibfnamefont {C.~M.}\ \bibnamefont
  {Warsop}}, \bibinfo {author} {\bibfnamefont {R.~E.}\ \bibnamefont
  {Williamson}}, \bibinfo {author} {\bibfnamefont {J.}~\bibnamefont
  {Pasternak}}, \bibinfo {author} {\bibfnamefont {J.~K.}\ \bibnamefont
  {Pozimski}}, \bibinfo {author} {\bibfnamefont {C.}~\bibnamefont {Brown}},\
  and\ \bibinfo {author} {\bibfnamefont {G.~H.}\ \bibnamefont {Rees}},\
  }\bibfield  {title} {\bibinfo {title} {{Progress on design studies for the
  \text{ISIS II} upgrade}},\ }in\ \href@noop {} {\emph {\bibinfo {booktitle}
  {Proc. 10th Int. Particle Accelerator Conf.}}}\ (\bibinfo {year} {2019})\ p.\
  \bibinfo {pages} {2075}\BibitemShut {NoStop}%
\bibitem [{\citenamefont {Bovet}\ \emph {et~al.}(1970)\citenamefont {Bovet},
  \citenamefont {Gouiran}, \citenamefont {Gumowski},\ and\ \citenamefont
  {Reich}}]{spacechargetuneshift}%
  \BibitemOpen
  \bibfield  {author} {\bibinfo {author} {\bibfnamefont {C.}~\bibnamefont
  {Bovet}}, \bibinfo {author} {\bibfnamefont {R.}~\bibnamefont {Gouiran}},
  \bibinfo {author} {\bibfnamefont {I.}~\bibnamefont {Gumowski}},\ and\
  \bibinfo {author} {\bibfnamefont {K.~H.}\ \bibnamefont {Reich}},\ }\bibfield
  {title} {\bibinfo {title} {A selection of formulae and data useful for the
  design of \text{A.G.} synchrotrons},\ }\href@noop {} {\bibfield  {journal}
  {\bibinfo  {journal} {CERN/MPS-SI/Int. DL/70/4}\ } (\bibinfo {year}
  {1970})}\BibitemShut {NoStop}%
\bibitem [{\citenamefont {Okawa}(1953)}]{okawaffa}%
  \BibitemOpen
  \bibfield  {author} {\bibinfo {author} {\bibfnamefont {T.}~\bibnamefont
  {Okawa}},\ }in\ \href@noop {} {\emph {\bibinfo {booktitle} {Proc. Annual
  meeting of \text{JPS} (unpublished)}}}\ (\bibinfo {year} {1953})\BibitemShut
  {NoStop}%
\bibitem [{\citenamefont {Symon}\ \emph {et~al.}(1956)\citenamefont {Symon},
  \citenamefont {Kerst}, \citenamefont {Jones}, \citenamefont {L.J.~Laslett},\
  and\ \citenamefont {Terwillinger}}]{ffa_d}%
  \BibitemOpen
  \bibfield  {author} {\bibinfo {author} {\bibfnamefont {K.}~\bibnamefont
  {Symon}}, \bibinfo {author} {\bibfnamefont {D.~W.}\ \bibnamefont {Kerst}},
  \bibinfo {author} {\bibfnamefont {L.~W.}\ \bibnamefont {Jones}}, \bibinfo
  {author} {\bibfnamefont {L.~J.}\ \bibnamefont {L.J.~Laslett}},\ and\ \bibinfo
  {author} {\bibfnamefont {K.~M.}\ \bibnamefont {Terwillinger}},\ }\bibfield
  {title} {\bibinfo {title} {{Fixed-Field Alternating-Gradient Particle
  Accelerators}},\ }\href@noop {} {\bibfield  {journal} {\bibinfo  {journal}
  {Phys. Rev.}\ }\textbf {\bibinfo {volume} {103}},\ \bibinfo {pages} {1837}
  (\bibinfo {year} {1956})}\BibitemShut {NoStop}%
\bibitem [{\citenamefont {Kolomensky}\ and\ \citenamefont
  {Lebedev}(1966{\natexlab{a}})}]{ffa_e}%
  \BibitemOpen
  \bibfield  {author} {\bibinfo {author} {\bibfnamefont {A.~A.}\ \bibnamefont
  {Kolomensky}}\ and\ \bibinfo {author} {\bibfnamefont {A.~N.}\ \bibnamefont
  {Lebedev}},\ }\href@noop {} {\emph {\bibinfo {title} {{Theory of Cyclic
  Accelerators}}}}\ (\bibinfo  {publisher} {North-Holland, Amsterdam},\
  \bibinfo {year} {1966})\BibitemShut {NoStop}%
\bibitem [{\citenamefont {Khoe}\ and\ \citenamefont {Kustom}(1983)}]{ffa_b}%
  \BibitemOpen
  \bibfield  {author} {\bibinfo {author} {\bibfnamefont {T.~K.}\ \bibnamefont
  {Khoe}}\ and\ \bibinfo {author} {\bibfnamefont {R.~L.}\ \bibnamefont
  {Kustom}},\ }\bibfield  {title} {\bibinfo {title} {{ASPUN} design for an
  argonne super intense pulsed neutron source},\ }\href@noop {} {\bibfield
  {journal} {\bibinfo  {journal} {IEEE Trans. on Nuclear Science}\ }\textbf
  {\bibinfo {volume} {NS-30}},\ \bibinfo {pages} {2086} (\bibinfo {year}
  {1983})}\BibitemShut {NoStop}%
\bibitem [{\citenamefont {Machida}(2017)}]{ffa_c}%
  \BibitemOpen
  \bibfield  {author} {\bibinfo {author} {\bibfnamefont {S.}~\bibnamefont
  {Machida}},\ }\bibfield  {title} {\bibinfo {title} {{Scaling Fixed-Field
  Alternating-Gradient Accelerators with Reverse Bend and Spiral Edge Angle}},\
  }\href@noop {} {\bibfield  {journal} {\bibinfo  {journal} {Phys. Rev. Lett.}\
  }\textbf {\bibinfo {volume} {119}},\ \bibinfo {pages} {064802} (\bibinfo
  {year} {2017})}\BibitemShut {NoStop}%
\bibitem [{\citenamefont {Aiba}\ \emph {et~al.}(2000)\citenamefont {Aiba},
  \citenamefont {Koba}, \citenamefont {Machida}, \citenamefont {Mori},
  \citenamefont {Muramatsu}, \citenamefont {Ohmori}, \citenamefont {Sakai},
  \citenamefont {Sato}, \citenamefont {Takagi}, \citenamefont {Ueno},
  \citenamefont {Yokoi}, \citenamefont {Yoshimoto},\ and\ \citenamefont
  {Yuasa}}]{popffa}%
  \BibitemOpen
  \bibfield  {author} {\bibinfo {author} {\bibfnamefont {M.}~\bibnamefont
  {Aiba}}, \bibinfo {author} {\bibfnamefont {K.}~\bibnamefont {Koba}}, \bibinfo
  {author} {\bibfnamefont {S.}~\bibnamefont {Machida}}, \bibinfo {author}
  {\bibfnamefont {Y.}~\bibnamefont {Mori}}, \bibinfo {author} {\bibfnamefont
  {R.}~\bibnamefont {Muramatsu}}, \bibinfo {author} {\bibfnamefont
  {C.}~\bibnamefont {Ohmori}}, \bibinfo {author} {\bibfnamefont
  {I.}~\bibnamefont {Sakai}}, \bibinfo {author} {\bibfnamefont
  {Y.}~\bibnamefont {Sato}}, \bibinfo {author} {\bibfnamefont {A.}~\bibnamefont
  {Takagi}}, \bibinfo {author} {\bibfnamefont {R.}~\bibnamefont {Ueno}},
  \bibinfo {author} {\bibfnamefont {T.}~\bibnamefont {Yokoi}}, \bibinfo
  {author} {\bibfnamefont {M.}~\bibnamefont {Yoshimoto}},\ and\ \bibinfo
  {author} {\bibfnamefont {Y.}~\bibnamefont {Yuasa}},\ }\bibfield  {title}
  {\bibinfo {title} {{Development of a FFAG proton synchrotron}},\ }in\
  \href@noop {} {\emph {\bibinfo {booktitle} {Proc. of EPAC 2000}}}\ (\bibinfo
  {year} {2000})\ p.\ \bibinfo {pages} {581}\BibitemShut {NoStop}%
\bibitem [{\citenamefont {Thomason}(2019)}]{thomason_isis_2019}%
  \BibitemOpen
  \bibfield  {author} {\bibinfo {author} {\bibfnamefont {J.}~\bibnamefont
  {Thomason}},\ }\bibfield  {title} {\bibinfo {title} {The {ISIS} {Spallation}
  {Neutron} and {Muon} {Source}--{The} {first thirty-three years}},\ }\bibfield
   {journal} {\bibinfo  {journal} {Nuclear Instruments and Methods in Physics
  Research Section A: Accelerators, Spectrometers, Detectors and Associated
  Equipment}\ }\href {https://doi.org/10.1016/j.nima.2018.11.129}
  {10.1016/j.nima.2018.11.129} (\bibinfo {year} {2019})\BibitemShut {NoStop}%
\bibitem [{\citenamefont {Terwilliger}\ \emph {et~al.}(1957)\citenamefont
  {Terwilliger}, \citenamefont {Jones},\ and\ \citenamefont
  {Pruett}}]{BeamStackingRSI}%
  \BibitemOpen
  \bibfield  {author} {\bibinfo {author} {\bibfnamefont {K.~M.}\ \bibnamefont
  {Terwilliger}}, \bibinfo {author} {\bibfnamefont {L.~W.}\ \bibnamefont
  {Jones}},\ and\ \bibinfo {author} {\bibfnamefont {C.~H.}\ \bibnamefont
  {Pruett}},\ }\bibfield  {title} {\bibinfo {title} {Beam stacking experiments
  in an electron model {FFAG} accelerator},\ }\href@noop {} {\bibfield
  {journal} {\bibinfo  {journal} {Review of Scientific Instruments}\ }\textbf
  {\bibinfo {volume} {28}},\ \bibinfo {pages} {987} (\bibinfo {year}
  {1957})}\BibitemShut {NoStop}%
\bibitem [{\citenamefont {Myers}(2016)}]{BeamStackingISR}%
  \BibitemOpen
  \bibfield  {author} {\bibinfo {author} {\bibfnamefont {S.}~\bibnamefont
  {Myers}},\ }\href {https://doi.org/10.1142/9789814436403_0009} {\emph
  {\bibinfo {title} {Challenges and goals for accelerators in the \text{XXI}
  century}}}\ (\bibinfo  {publisher} {World Scientific Publishing Co Pte Ltd},\
  \bibinfo {year} {2016})\ Chap.\ \bibinfo {chapter} {9, The {CERN Intersecting
  Storage Rings}}\BibitemShut {NoStop}%
\bibitem [{\citenamefont {Pyeon}\ and\ \citenamefont {et. al.}(2009)}]{pyeon}%
  \BibitemOpen
  \bibfield  {author} {\bibinfo {author} {\bibfnamefont {C.~H.}\ \bibnamefont
  {Pyeon}}\ and\ \bibinfo {author} {\bibnamefont {et. al.}},\ }\bibfield
  {title} {\bibinfo {title} {First injection of spallation neutrons generated
  by high-energy protons into the kyoto university critical assembly},\
  }\href@noop {} {\bibfield  {journal} {\bibinfo  {journal} {Journal of Nuclear
  Science and Technology}\ }\textbf {\bibinfo {volume} {46}},\ \bibinfo {pages}
  {1091} (\bibinfo {year} {2009})}\BibitemShut {NoStop}%
\bibitem [{\citenamefont {Sheehy}\ \emph {et~al.}(2016)\citenamefont {Sheehy},
  \citenamefont {Kelliher}, \citenamefont {Machida}, \citenamefont {Rogers},
  \citenamefont {Prior}, \citenamefont {Volat}, \citenamefont {Haj~Tahar},
  \citenamefont {Ishi}, \citenamefont {Kuriyama}, \citenamefont {Sakamoto},
  \citenamefont {Uesugi},\ and\ \citenamefont {Mori}}]{KURRIFFA}%
  \BibitemOpen
  \bibfield  {author} {\bibinfo {author} {\bibfnamefont {S.~L.}\ \bibnamefont
  {Sheehy}}, \bibinfo {author} {\bibfnamefont {D.~J.}\ \bibnamefont
  {Kelliher}}, \bibinfo {author} {\bibfnamefont {S.}~\bibnamefont {Machida}},
  \bibinfo {author} {\bibfnamefont {C.}~\bibnamefont {Rogers}}, \bibinfo
  {author} {\bibfnamefont {C.~R.}\ \bibnamefont {Prior}}, \bibinfo {author}
  {\bibfnamefont {L.}~\bibnamefont {Volat}}, \bibinfo {author} {\bibfnamefont
  {M.}~\bibnamefont {Haj~Tahar}}, \bibinfo {author} {\bibfnamefont
  {Y.}~\bibnamefont {Ishi}}, \bibinfo {author} {\bibfnamefont {Y.}~\bibnamefont
  {Kuriyama}}, \bibinfo {author} {\bibfnamefont {M.}~\bibnamefont {Sakamoto}},
  \bibinfo {author} {\bibfnamefont {T.}~\bibnamefont {Uesugi}},\ and\ \bibinfo
  {author} {\bibfnamefont {Y.}~\bibnamefont {Mori}},\ }\bibfield  {title}
  {\bibinfo {title} {{Characterization techniques for fixed-field alternating
  gradient accelerators and beam studies using the \text{KURRI} 150 \text{MeV}
  proton FFAG}},\ }\bibfield  {journal} {\bibinfo  {journal} {Prog. Theor. Exp.
  Phys.}\ }\textbf {\bibinfo {volume} {073G01}},\ \href
  {https://doi.org/10.1093/ptep/ptw086} {10.1093/ptep/ptw086} (\bibinfo {year}
  {2016})\BibitemShut {NoStop}%
\bibitem [{\citenamefont {Uesugi}\ and\ \citenamefont {et.
  al.}(2011)}]{tom_injection}%
  \BibitemOpen
  \bibfield  {author} {\bibinfo {author} {\bibfnamefont {T.}~\bibnamefont
  {Uesugi}}\ and\ \bibinfo {author} {\bibnamefont {et. al.}},\ }\bibfield
  {title} {\bibinfo {title} {{RF capture of Beam with Charge-Exchanging
  Multi-Turn Injection}},\ }in\ \href@noop {} {\emph {\bibinfo {booktitle}
  {Proc. 2nd Int. Particle Accelerator Conf.}}}\ (\bibinfo {year} {2011})\ p.\
  \bibinfo {pages} {454}\BibitemShut {NoStop}%
\bibitem [{\citenamefont {Kolomensky}\ and\ \citenamefont
  {Lebedev}(1966{\natexlab{b}})}]{PhaseDisplacement}%
  \BibitemOpen
  \bibfield  {author} {\bibinfo {author} {\bibfnamefont {A.~A.}\ \bibnamefont
  {Kolomensky}}\ and\ \bibinfo {author} {\bibfnamefont {A.~N.}\ \bibnamefont
  {Lebedev}},\ }in\ \href@noop {} {\emph {\bibinfo {booktitle} {Theory of
  cyclic accelerators}}}\ (\bibinfo  {publisher} {North-Holland, Amsterdam},\
  \bibinfo {year} {1966})\ Chap.\ \bibinfo {chapter} {{7, Peculiarities of
  accelerators of different types}}\BibitemShut {NoStop}%
\bibitem [{\citenamefont {Symon}\ and\ \citenamefont
  {Sessler}(1956)}]{Symon_Sessler}%
  \BibitemOpen
  \bibfield  {author} {\bibinfo {author} {\bibfnamefont {K.~R.}\ \bibnamefont
  {Symon}}\ and\ \bibinfo {author} {\bibfnamefont {A.~M.}\ \bibnamefont
  {Sessler}},\ }\bibfield  {title} {\bibinfo {title} {{Methods of Radio
  Frequency Acceleration in Fixed Field Accelerators with Applications to High
  Current and Intersecting Beam Accelerators}},\ }in\ \href
  {https://cds.cern.ch/record/1052548/files/p44.pdf} {\emph {\bibinfo
  {booktitle} {{Conf.Proc.C }}}},\ \bibinfo {series} {Heacc 56}, Vol.\ \bibinfo
  {volume} {560611}\ (\bibinfo {year} {1956})\BibitemShut {NoStop}%
\bibitem [{\citenamefont {Jones}\ \emph {et~al.}(1959)\citenamefont {Jones},
  \citenamefont {Pruett}, \citenamefont {Symon},\ and\ \citenamefont
  {Terwilliger}}]{Jones}%
  \BibitemOpen
  \bibfield  {author} {\bibinfo {author} {\bibfnamefont {L.~W.}\ \bibnamefont
  {Jones}}, \bibinfo {author} {\bibfnamefont {C.~H.}\ \bibnamefont {Pruett}},
  \bibinfo {author} {\bibfnamefont {K.~R.}\ \bibnamefont {Symon}},\ and\
  \bibinfo {author} {\bibfnamefont {K.~M.}\ \bibnamefont {Terwilliger}},\
  }\bibfield  {title} {\bibinfo {title} {{Comparison of Experimental Results
  with the Theory of Radio-Frequency Acceleration Processes in FFAG
  Accelerators}},\ }in\ \href@noop {} {\emph {\bibinfo {booktitle} {2nd
  International Conference on High-Energy Accelerators}}}\ (\bibinfo {year}
  {1959})\ p.~\bibinfo {pages} {58}\BibitemShut {NoStop}%
\bibitem [{\citenamefont {Turner}(1987)}]{CASschottky}%
  \BibitemOpen
  \bibfield  {author} {\bibinfo {author} {\bibfnamefont {S.}~\bibnamefont
  {Turner}},\ }\href {https://doi.org/10.5170/CERN-1987-003-V-2} {\emph
  {\bibinfo {title} {{CAS - CERN Advanced Accelerator School}}}}\ (\bibinfo
  {publisher} {CERN},\ \bibinfo {year} {1987})\BibitemShut {NoStop}%
\bibitem [{\citenamefont {Nolden}(2001)}]{nolden_instrumentation_2001}%
  \BibitemOpen
  \bibfield  {author} {\bibinfo {author} {\bibfnamefont {F.}~\bibnamefont
  {Nolden}},\ }\bibfield  {title} {\bibinfo {title} {{Instrumentation And
  Diagnostics Using Schottky Signals}},\ }in\ \href
  {https://cds.cern.ch/record/923649} {\emph {\bibinfo {booktitle} {{Dipac}}}}\
  (\bibinfo {year} {2001})\BibitemShut {NoStop}%
\bibitem [{\citenamefont {Stoica}\ and\ \citenamefont
  {Moses}(2005)}]{petre_stoica_spectral_2005}%
  \BibitemOpen
  \bibfield  {author} {\bibinfo {author} {\bibfnamefont {P.}~\bibnamefont
  {Stoica}}\ and\ \bibinfo {author} {\bibfnamefont {R.~L.}\ \bibnamefont
  {Moses}},\ }\href@noop {} {\emph {\bibinfo {title} {{Spectral Analysis of
  Signals}}}}\ (\bibinfo  {publisher} {Prentice Hall},\ \bibinfo {year}
  {2005})\BibitemShut {NoStop}%
\bibitem [{\citenamefont {Bartlett}(1950)}]{bartlett_periodogram_1950}%
  \BibitemOpen
  \bibfield  {author} {\bibinfo {author} {\bibfnamefont {M.~S.}\ \bibnamefont
  {Bartlett}},\ }\bibfield  {title} {\bibinfo {title} {{Periodogram Analysis
  and Continuous Spectra}},\ }\href@noop {} {\bibfield  {journal} {\bibinfo
  {journal} {Biometrika}\ }\textbf {\bibinfo {volume} {37}} (\bibinfo {year}
  {1950})}\BibitemShut {NoStop}%
\bibitem [{\citenamefont {Welch}(1967)}]{welch_use_1967}%
  \BibitemOpen
  \bibfield  {author} {\bibinfo {author} {\bibfnamefont {P.}~\bibnamefont
  {Welch}},\ }\bibfield  {title} {\bibinfo {title} {{The use of Fast Fourier
  Transform for the Estimation of Power Spectra}},\ }\bibfield  {journal}
  {\bibinfo  {journal} {{IEEE} Transactions on Audio and Electroacoustics}\
  }\textbf {\bibinfo {volume} {15}},\ \href
  {https://doi.org/10.1109/tau.1967.1161901} {10.1109/tau.1967.1161901}
  (\bibinfo {year} {1967})\BibitemShut {NoStop}%
\bibitem [{\citenamefont {Uesugi}\ \emph {et~al.}(2019)\citenamefont {Uesugi},
  \citenamefont {Fuwa}, \citenamefont {Ishi}, \citenamefont {Kuriyama},\ and\
  \citenamefont {Mori}}]{fab_ipac}%
  \BibitemOpen
  \bibfield  {author} {\bibinfo {author} {\bibfnamefont {T.}~\bibnamefont
  {Uesugi}}, \bibinfo {author} {\bibfnamefont {Y.}~\bibnamefont {Fuwa}},
  \bibinfo {author} {\bibfnamefont {Y.}~\bibnamefont {Ishi}}, \bibinfo {author}
  {\bibfnamefont {Y.}~\bibnamefont {Kuriyama}},\ and\ \bibinfo {author}
  {\bibfnamefont {Y.}~\bibnamefont {Mori}},\ }\bibfield  {title} {\bibinfo
  {title} {{Study of Beam Injection Efficiency in the Fixed Field Alternating
  Gradient Synchrotron in KURNS}},\ }in\ \href@noop {} {\emph {\bibinfo
  {booktitle} {Proc. 10th Int. Particle Accelerator Conf.}}}\ (\bibinfo {year}
  {2019})\ p.\ \bibinfo {pages} {2564}\BibitemShut {NoStop}%
\bibitem [{\citenamefont {Terwilliger}(1956)}]{terwilliger_radio_1956}%
  \BibitemOpen
  \bibfield  {author} {\bibinfo {author} {\bibfnamefont {K.~M.}\ \bibnamefont
  {Terwilliger}},\ }\bibfield  {title} {\bibinfo {title} {Radio frequency
  knockout of stacked beams},\ }\href@noop {} {\bibfield  {journal} {\bibinfo
  {journal} {MURA-133, MURA-KMT-3}\ } (\bibinfo {year} {1956})}\BibitemShut
  {NoStop}%
\bibitem [{\citenamefont {Jones}\ \emph {et~al.}(1957)\citenamefont {Jones},
  \citenamefont {Pruett},\ and\ \citenamefont
  {Terwilliger}}]{jones_experiments_1957}%
  \BibitemOpen
  \bibfield  {author} {\bibinfo {author} {\bibfnamefont {L.~W.}\ \bibnamefont
  {Jones}}, \bibinfo {author} {\bibfnamefont {C.~H.}\ \bibnamefont {Pruett}},\
  and\ \bibinfo {author} {\bibfnamefont {K.~M.}\ \bibnamefont {Terwilliger}},\
  }\bibfield  {title} {\bibinfo {title} {Experiments on radio frequency
  knockout of stacked beams},\ }\href@noop {} {\bibfield  {journal} {\bibinfo
  {journal} {MURA-260}\ } (\bibinfo {year} {1957})}\BibitemShut {NoStop}%
\bibitem [{\citenamefont {Kerst}(1957)}]{kerst_suppression_1957}%
  \BibitemOpen
  \bibfield  {author} {\bibinfo {author} {\bibfnamefont {D.~W.}\ \bibnamefont
  {Kerst}},\ }\bibfield  {title} {\bibinfo {title} {Suppression of betatron
  oscillation excitation ({RF} knockout) by the {RF} accelerating system of a
  fixed field accelerator},\ }\href@noop {} {\bibfield  {journal} {\bibinfo
  {journal} {MURA-215}\ } (\bibinfo {year} {1957})}\BibitemShut {NoStop}%
\end{thebibliography}%

\end{document}